\documentclass[12pt]{article}

\usepackage{amsmath,amssymb,cite}
\usepackage{graphicx}

\makeatletter
 
  \@addtoreset{equation}{section}
 \makeatother

\newcommand{\be}{\begin{equation}}
\newcommand{\ee}{\end{equation}}
\newcommand{\bea}{\begin{eqnarray}}
\newcommand{\eea}{\end{eqnarray}}
\newcommand{\beann}{\begin{eqnarray*}}
\newcommand{\eeann}{\end{eqnarray*}}

\newcommand{\ba}{\begin{array}}
\newcommand{\ea}{\end{array}}
\newcommand{\dif}[1]{{d}{#1}}

\newcommand{\tr}{\mbox{tr}}
\newcommand{\Tr}{\mathop{\rm Tr}}

\newcommand{\IN}{\mathbb{N}}

\newcommand{\non}{\nonumber}
\newcommand{\s}{{\sigma}}
\newcommand{\al}{\alpha}
\newcommand{\halfL}{\frac{\Lambda}{2}}
\graphicspath{{./graph/}}
\usepackage{subfigure,comment}

\newcommand{\n}{\nonumber \\}

\def\Xint#1{\mathchoice
   {\XXint\displaystyle\textstyle{#1}}%
   {\XXint\textstyle\scriptstyle{#1}}%
   {\XXint\scriptstyle\scriptscriptstyle{#1}}%
   {\XXint\scriptscriptstyle\scriptscriptstyle{#1}}%
   \!\int}
\def\XXint#1#2#3{{\setbox0=\hbox{$#1{#2#3}{\int}$}
     \vcenter{\hbox{$#2#3$}}\kern-.5\wd0}}

\def\dashint{\Xint-}

\allowdisplaybreaks

\begin{document}

\setlength{\oddsidemargin}{0cm}
\setlength{\baselineskip}{7mm}

\begin{titlepage}
\renewcommand{\thefootnote}{\fnsymbol{footnote}}
\begin{normalsize}
\begin{flushright}
\begin{tabular}{l}
KUNS-2387
\end{tabular}
\end{flushright}
  \end{normalsize}

~~\\

\vspace*{0cm}
    \begin{Large}
       \begin{center}
         {
          Large-$N$ reduction for 
         ${\cal N}=2$ quiver Chern-Simons theories on $S^3$\\
 and localization in matrix models \\ }
       \end{center}
    \end{Large}
\vspace{0.7cm}

\begin{center}
Yuhma A{\sc sano}\footnote
            {
e-mail address : 
yuhma@gauge.scphys.kyoto-u.ac.jp }, 
Goro I{\sc shiki}\footnote
            {
e-mail address : 
ishiki@post.kek.jp}, 
Takashi O{\sc kada}\footnote
            {
e-mail address : 
okada@gauge.scphys.kyoto-u.ac.jp }
    {\sc and}
Shinji S{\sc himasaki}\footnote
           {
e-mail address : 
shinji@gauge.scphys.kyoto-u.ac.jp }\\
      
\vspace{0.7cm}
                    
      {\it Department of Physics, Kyoto University}\\
               {\it Kyoto, 606-8502, Japan}\\
          
\end{center}

\vspace{0.7cm}

\begin{abstract}
\noindent
We study reduced matrix models obtained by 
the dimensional reduction of ${\cal N}=2$ quiver
Chern-Simons theories on $S^3$ to zero dimension and show that 
if a reduced model is expanded
around a particular multiple fuzzy sphere background, 
it becomes equivalent to the original theory on $S^3$
in the large-$N$ limit. 
This is regarded as a novel large-$N$ reduction on a curved 
space $S^3$.
We perform the localization method to the reduced model
and compute the free energy and the vacuum expectation value of 
a BPS Wilson loop operator.
In the large-$N$ limit, we find 
an exact agreement between these results and those in 
the original theory on $S^3$.

\end{abstract}
\vfill
\end{titlepage}
\vfil\eject

\setcounter{footnote}{0}

\tableofcontents

\section{Introduction}
The localization technique has attracted much attention in 
recent years as an efficient method of computing
a class of physical quantities of our interest.
For instance, it enables us to compute exactly BPS Wilson loops
or partition functions in 4d ${\cal N}=2$ 
supersymmetric Yang-Mills (SYM) theories \cite{Pestun:2007rz}
or those in 3d ${\cal N}=2$ Chern-Simons (CS) theories 
coupled to some matter fields
\cite{Kapustin:2009kz,Drukker:2010nc,Jafferis:2010un,Hama:2010av,Drukker:2011zy,Fuji:2011km}.
In particular, predictions from their gravity duals, such as 
the $N^{3/2}$ law of the free energy \cite{Klebanov:1996un}
in the ABJM theory\cite{Aharony:2008ug}, 
can be verified explicitly based on the localization method 
\cite{Drukker:2010nc} and thus
a remarkable progress has been made in testing the 
gauge/gravity correspondence
\cite{Maldacena:1997re}.

In this paper, we use the localization method for another purpose.
We apply it to a dimensionally reduced matrix model of 
a general ${\cal N}=2$ non-chiral\footnote{
In quiver diagram, non-chiral means that for every arrow from node A to node B 
there is a corresponding arrow from node B to node A.
} quiver CS theory on $S^3$ and show that there exists 
an equivalence in a large-$N$ limit between the reduced 
model and the original theory on $S^3$. This kind of 
large-$N$ equivalence on $S^3$ was first discovered in 
\cite{Ishii:2008ib} for 
${\cal N}=4$ SYM on $R\times S^3$ (see also
\cite{Ishii:2008tm,Ishii:2007ex,Ishiki:2006yr} for earlier discussions)
and it is regarded as 
a novel type of the large-$N$ reduction extended to the case of 
$S^3$.

The original large-$N$ reduction initiated by Eguchi and Kawai 
asserts that a gauge theory on flat space-time in 
the planar limit is equivalent to a matrix model 
that is obtained by its dimensional 
reduction to lower dimensions \cite{Eguchi:1982nm}.
This equivalence is significant not only because 
it realizes the emergence of space-time in matrix models, 
which is relevant in the context of the 
matrix model formulation for string theories \cite{BFSS},
but also because this equivalence implies that the matrix models provide
a non-perturbative formulation of planar gauge theories
which is alternative to the lattice formulation.
One may expect that supersymmetric theories can be described in terms 
of matrix models non-perturbatively based on this equivalence 
while it is generally difficult in the lattice formulation\footnote{
There are considerable recent developments in the lattice
theories for supersymmetric theories \cite{Kaplan:2005ta}.}.

It is however well-known that this equivalence fails to hold due 
to the spontaneous $U(1)^D$ symmetry breaking in the original 
Eguchi-Kawai model \cite{Bhanot:1982sh}. 
To overcome this difficulty, the quenching and the twisting 
prescriptions were proposed 
\cite{Bhanot:1982sh,Parisi:1982gp,Gross:1982at,Das:1982ux,
GonzalezArroyo:1982hz}. 
Although the symmetry breaking can be avoided by introducing such prescriptions 
at least when the theory has a sufficient number of fermions\footnote{
See \cite{Kovtun:2007py,Azeyanagi:2010ne} for detail.
},
they do not preserve supersymmetry.
Because of this, it had been difficult until recently to construct a 
non-perturbative formulation of supersymmetric gauge theories based on 
the large-$N$ reduction which keeps supersymmetry manifestly.

The novel large-$N$ reduction was proposed
for theories on $S^3$ \cite{Ishii:2008ib}. 
It states that a reduced model, which is obtained by 
the dimensional reduction of a gauge theory on $S^3$ to a point,
becomes equivalent to the original gauge theory 
in the large-$N$ limit if the reduced model 
is expanded around a certain multiple 
fuzzy sphere background and a continuum limit is taken.
In this proposal, the above-mentioned difficulty
is avoided thanks to the curvature of $S^3$.
On $S^3$, a gauge theory acquires a mass gap, which is inversely 
proportional to the radius of $S^3$,
and hence does not possess 
a flat direction, which would lead the symmetry breaking and 
spoil the large-$N$ equivalence.
This implies that the prescriptions are not needed in this case,
so that a reduced model 
obtained from a supersymmetric gauge theory
still keeps part of the original supersymmetry. 
One can therefore use the reduced model 
as a non-perturbative formulation of the supersymmetric theory on $S^3$.
So far, such formulation has been considered for ${\cal N}=4$ SYM
\cite{Ishii:2008ib}, SYM with lower supersymmetry
\cite{Hanada:2009kz}
and supersymmetric quiver CS theories \cite{Hanada:2009hd}.
The large-$N$ reduction has also been extended to the cases for 
more general manifolds such as group manifolds \cite{Kawai:2009vb} 
and coset spaces \cite{Kawai:2010sf}.

In particular, the large-$N$ reduction for
${\cal N}=4$ SYM on $R\times S^3$ 
 has been studied actively 
\cite{Ishiki:2008te,Kitazawa:2008mx,Ishiki:2009sg,Ishiki:2011ct}
since it is relevant to testing
the original version of the AdS/CFT correspondence for the type IIB 
string theory on $AdS_5 \times S^5$.
The reduced model of ${\cal N}=4$ SYM on $R\times S^3$
is given by the plane wave matrix model (PWMM)
\cite{Berenstein:2002jq,Kim:2003rza}.
This model preserves $SU(2|4)$ symmetry,
which can not be realized in the lattice formulation at present,
so that it is expected to describe the original theory 
on $S^3$ in the continuum limit without any fine-tuning.
Based on this formulation, one can analyze numerically the strongly 
coupled regime of the planar ${\cal N}=4$ SYM, which is mapped to the 
regime in the string theory where the supergravity or 
the semiclassical approximation is valid.
The methods of the Monte Carlo simulation for matrix models 
proposed in 
\cite{Hanada:2007ti,Anagnostopoulos:2007fw,Catterall:2008yz} 
are available for the numerical computation.
Thus, it gives a feasible way of testing the AdS/CFT 
correspondence\footnote{See 
\cite{Honda:2011qk,Honda:2010nx} for preliminary results of such attempts.}.

Although the validity of such non-perturbative formulation 
of supersymmetric theories has been checked by some 
perturbative calculations
\cite{Ishiki:2008te,Kitazawa:2008mx,Ishiki:2009sg,Ishiki:2011ct}, 
it should be checked also for the strong coupling region.
In this paper, we consider the large-$N$ reduction for 
${\cal N}=2$ non-chiral quiver CS theories on $S^3$. 
Applying the localization method,
we compute the partition function and the one-point function of 
the great circular BPS Wilson loop operator in the reduced model.
Then we prove the large-$N$ equivalence for these quantities 
to all orders in the perturbation theory. 
We also find that a saddle point configuration of the reduced matrix model 
is given as infinitely many copies of that in the original theory 
up to a cutoff effect, which is negligible in the continuum limit.
This fact ensures that the equivalence also holds 
even in the strongly coupled regime. 

So far, a similar test of the large-$N$ reduction has been done 
for the pure CS theory on $S^3$ 
\cite{Ishiki:2009vr,Ishiki:2010pe}\footnote{See also 
\cite{Ishii:2007sy,IOST}.}, 
which is a solvable topological field theory \cite{Witten:1988hf}.
Since the path integral of 
the reduced model is easily performed in this case, 
it is possible to see the agreement of two theories
through a direct calculation.
Although the theory we consider in this paper contains dynamical 
degrees of freedom and hence is not so simple,
the localization method enables us to 
verify the large-$N$ equivalence explicitly.

This paper is organized as follows. 
In section 2, we review known results on the computation 
of the partition function and the BPS Wilson loop operator in 
a general ${\cal N}=2$ quiver CS theory.
In section 3, we introduce the reduced model of the theory on $S^3$
focusing on the supersymmetry transformation.
We also perform the path integral of the reduced model 
by means of the localization.
In section 4, we show the large-$N$ equivalence.
We first extract a theory around the multiple fuzzy sphere background
which creates $S^3$
from the result of the localization
and see that the theory reduces to a certain eigenvalue integral.
By studying this integral, we see the 
equivalence both in the perturbation theory and in the saddle point equation.
Section 5 is devoted to conclusion. In appendices, some details are gathered.

\section{Localization in ${\cal N}=2$ quiver CS theory on $S^3$}
In this section, we review some known results for the localization in
a ${\cal N}=2$ quiver CS theory on $S^3$ 
\cite{Kapustin:2009kz,Drukker:2010nc,Jafferis:2010un,Hama:2010av}.
We assume that the gauge group is given by a product of unitary groups,
$\bigotimes_{a}U(N_a)$, and consider a general matter content.
We set the radius of $S^3$ to be one in this paper and 
our convention for the theory on $S^3$ is 
summarized in appendix \ref{ApC}.

In this theory, there are nilpotent supersymmetries, which we will call $Q$ 
symbolically in the following.
The partition function is invariant under adding $Q$-exact terms 
to the action. Hence the path integral can be localized onto 
the saddle points of the $Q$-exact terms.
For a gauge multiplet, the role of the $Q$-exact term is played by the 
Yang-Mills (YM) action on $S^3$ and for a matter multiplet by a part of 
the matter action. 
The saddle point configuration is given by 
the flat connection for the gauge field, which is trivial on $S^3$, 
namely, $A_{\mu}=0$ up to gauge transformation. Also all the matter fields 
are zero at the saddle point.  The other bosonic fields in the vector 
multiplet, $\sigma$ and $D$, take 
nontrivial values at the saddle point. They are given by,
\begin{align}
\sigma=-D= {\rm constant}.
\label{2.1}
\end{align}
Then, the calculation of the partition function amounts to computing the 
1-loop determinant at each saddle point.
The result is written as a summation over contributions 
from all the saddle points, 
namely in our case, as an integral over the constant matrix $\sigma$ for 
each gauge multiplet.
In the following, we will work in the gauge in which $\sigma$'s in 
all the gauge multiplets are diagonalized,
so that the integration measure has the Vandermonde
determinant as $\int \prod_i d\sigma_i \prod_{i<j}(\sigma_i -\sigma_j)^2$
for each gauge multiplet.

The contribution from a vector multiplet is given by 
\begin{align}
\prod_{i<j} \left( \frac{\sinh(\pi(\sigma_i-\sigma_j))}{\pi(\sigma_i-\sigma_j)}
\right)^2,
\label{2.2}
\end{align}
up to an overall constant. Note that the denominator cancels the Vandermonde 
determinant.

Then, let us consider the contribution from a matter multiplet in the bifundamental representation which couples to two different gauge multiplets.
The determinant takes the following form,
\begin{align}
\prod_{n>0} \prod_{i,\alpha}
\frac{n+1-q+i(\sigma_i-\rho_{\alpha})}{n-1-q-i(\sigma_i-\rho_{\alpha})},
\label{2.3}
\end{align}
where $q$ is the dimension of the lowest components in the matter multiplet 
and $\rho$ is the counterpart of $\sigma$ in the second gauge multiplet.
The determinant for an adjoint matter multiplet 
can be obtained by putting $\rho_i = \sigma_i$ in (\ref{2.3}).
In particular, when $q=\frac{1}{2}$, it is simplified and given by
\begin{align}
\prod_{i<j}\frac{1}{\cosh(\pi(\sigma_i-\sigma_j))}.
\label{2.4}
\end{align}

For example, 
the ABJM theory with gauge group $U(N_1)_k \times U(N_2)_{-k}$ contains 
two vector multiplets and four matter chiral multiplets in the bifundamental 
representation.
The partition function is reduced through the above calculation to 
the so-called ABJM matrix model,
\begin{align}
\int \prod_i d\sigma_i \prod_{\alpha} d\rho_{\alpha}
\frac{\prod_{i<j}\sinh^2(\pi(\sigma_i-\sigma_j))
\prod_{\alpha<\beta}\sinh^2(\pi(\rho_{\alpha}-\rho_{\beta}))}
{\prod_{i,\alpha}\cosh^2(\pi(\sigma_i-\rho_{\alpha}))}
e^{-\frac{2\pi^2}{g_s}\sum_{i}\sigma^2_i+\frac{2\pi^2}{g_s}
\sum_{\alpha}\rho^2_{\alpha}},
\label{2.5}
\end{align}
where the Gaussian factors are obtained by substituting the saddle point 
configuration to the original CS actions and the coupling constant 
is related to the CS level as $g_s= 2\pi i/k$.

A correlation function of $Q$-closed operators can also be 
reduced to an eigenvalue integral by the localization method.
We consider the one-point function of the BPS Wilson loop,
\begin{align}
W({\cal C})= \frac{1}{N} {\rm tr} P
\exp \left( {i\int^1_0 ds(\dot{x}^{\mu}(s)A_{\mu}(x)-i|\dot{x}(s)|\sigma (x))} 
\right),
\label{2.6}
\end{align}
where ${\rm tr}$ stands for the trace in the fundamental representation 
and $N$ is the rank of the gauge group for $A_{\mu}$.
We consider a great circle on $S^3$ as the contour ${\cal C}$.
It is parametrized as (see appendix \ref{s3} 
for our notation for $S^3$)
\begin{align}
\{x^{\mu}(s)\}=(\theta(s),\varphi(s),\psi(s))=(0,0,4\pi s),
\label{2.7}
\end{align}
with $s \in [0,1]$. In this case, the operator is BPS and $Q$-closed.
Evaluating the one-point function around the saddle point, 
we arrive at 
\begin{align}
W({\cal C})= \frac{1}{N} 
\sum_{i}
\left\langle 
e^{2\pi \sigma_i}
\right\rangle,
\label{2.8}
\end{align}
where $\langle \cdots \rangle $ stands for an average taken with
respect to the eigenvalue integral obtained by the 
localization of the partition function.
Note that in the large-$N$ limit, which is our main interest
in this paper, a general correlation function of the Wilson loops 
decomposes to a product of the one-point functions because of the 
factorization property.

The remaining task of this calculation would be 
to perform the eigenvalue integral.
Although there are several efficient ways of evaluating the integral
\cite{Drukker:2010nc, Marino:2011eh,Marino:2011nm,Fuji:2011km,Hanada:2012si}, 
we do not review them here since any explicit solution 
is not needed in this paper.
In the following sections, we consider the reduced model of 
the quiver CS theory and show that its partition function and 
the one-point function of a corresponding operator are equivalent 
to the above eigenvalue integrals of the theory on $S^3$
in the large-$N$ limit.

\section{Reduced model for ${\cal N}=2$ quiver CS theory on $S^3$}
In this section, we construct the reduced model of the ${\cal N}=2$ quiver
CS theory on $S^3$ and apply the localization calculation to the 
model.
We first perform the dimensional reduction from $S^3$ to 
a point to obtain the reduced model. 
Then, we perform the localization for each multiplet of the 
${\cal N}=2$ supersymmetry.
We list the action and the supersymmetry 
transformations in the original theory on $S^3$ in appendix \ref{ApC}.

\subsection{Dimensional reduction}
Let us first demonstrate the dimensional reduction of the CS term 
(\ref{scs}) for a single gauge multiplet.
In order to reduce it to a point,
we expand the gauge field 
$A$ in terms of the right-invariant one-form $e^a$ 
defined in (\ref{ri 1-form}) as,
\begin{equation}
A=A_\mu (x) dx^\mu=A_a(x)e^{a},
\end{equation}
and drop the coordinate dependence of $A_a$ and of 
the other fields in the multiplet. Using the 
Maurer-Cartan equation (\ref{Maurer-Cartan}), 
the derivative of $A$ can be calculated as follows,
\begin{align}
dA \rightarrow A_a de^{a} = \varepsilon^{abc} A_a e^{b}\wedge e^{c},
\label{dim red}
\end{align}
where the arrow represents that we have dropped the derivative of $A_a$.
By applying (\ref{dim red}) to (\ref{scs}) 
and performing the integral, which produces only an 
overall constant factor given by the volume of unit $S^3$, 
we obtain the reduced model of the CS term,
\begin{equation}
S^r_{CS}=-\frac{1}{g^2}\Tr \left[ A_aA^a-\frac{i}{3}\varepsilon^{abc}A_aA_bA_c-
\frac{1}{2}\bar \lambda \lambda +D\sigma \right] .
\label{SCSaction}
\end{equation}
Here, the indices $a,b,c$ are raised and lowered simply by the 
Kronecker delta and we have introduced a coupling 
constant $g$ for the reduced model. 
The value of $g$ will be determined later such that 
the original continuum theory is reproduced in a continuum limit.

The original theory on $S^3$ has two kinds of the Killing spinors
in the right-invariant frame. 
One is constant and the other is dependent on the coordinates of $S^3$
as shown in (\ref{solution to KSE}).
Under the dimensional reduction, the constant Killing spinors survive and 
they generate the supersymmetry transformations of the reduced model.
By dimensionally reducing the supersymmetry transformations in the original 
theory (\ref{susy for gauge on s3}), 
we find that (\ref{SCSaction}) is invariant under 
the following supersymmetry transformations:
\begin{align}
\delta A_a=&\frac{i}{2}(\bar \lambda \gamma _a\epsilon -\bar \epsilon \gamma _a\lambda ),\nonumber\\
\delta \sigma =&-\frac{1}{2}(\bar \lambda \epsilon -\bar \epsilon \lambda ),
\nonumber\\
\delta \lambda =&\frac{1}{2}\gamma ^{ab}\epsilon F_{ab}-D\epsilon +\gamma ^a\epsilon [A_a,\sigma ]-\sigma \epsilon ,
\nonumber\\
\delta \bar \lambda =&\frac{1}{2}\gamma ^{ab}\bar \epsilon F_{ab}+D\bar \epsilon -\gamma ^a\bar \epsilon [A_a,\sigma ]+\sigma \bar \epsilon ,
\nonumber\\
\delta D=&-\frac{1}{2}[A_a,\bar \lambda ]\gamma ^a \epsilon -\frac{1}{2}\bar 
\epsilon \gamma ^a [A_a,\lambda ]+\frac{i}{2}[\bar \lambda \epsilon ,\sigma ]
+\frac{i}{2}[\bar \epsilon \lambda ,\sigma ]
-\frac{1}{2}\bar \lambda \epsilon +\frac{1}{2}\bar \epsilon \lambda ,
\label{susygauge}
\end{align}
where
\begin{equation}
F_{ab}:=2\varepsilon _{abc}A^c-i[A_a,A_b].
\end{equation}
The transformations (\ref{susygauge}) 
consist of two independent parts 
generated by $\epsilon$ and $\bar{\epsilon}$ as 
$\delta= \delta_{\epsilon}+\delta_{\bar{\epsilon}}$.
The Killing spinors $\epsilon$ and $\bar{\epsilon}$ are 
constant complex two-component spinors, 
and decomposed into the upper and the lower components.
The transformation generated by each component is nilpotent.  
To see this, we take two parameters $\epsilon$ and $\epsilon'$ which 
have only the upper components and hence satisfy 
$\epsilon^\prime \epsilon =\epsilon_1^\prime \epsilon_2+\epsilon_2^\prime \epsilon_1=0$.
Then, we can see that the supersymmetry is nilpotent, 
$\delta _{\epsilon^\prime}\delta _{\epsilon}=0$.
For example, these transformations act on $\lambda$ as follows,
\begin{align}
\delta _{\epsilon ^\prime} \delta _{\epsilon}\lambda =&\frac{1}{2}\gamma ^{ab}\epsilon ( i\varepsilon _{abc}\bar \lambda \gamma ^c\epsilon ^\prime +[\bar \lambda \gamma _a\epsilon ^\prime ,A_b]) -\epsilon (-\frac{1}{2}[A_a,\bar \lambda \gamma ^a\epsilon ^\prime ]+\frac{i}{2}[\bar \lambda \epsilon ^\prime ,\sigma ]-\frac{1}{2}\bar \lambda \epsilon ^\prime )\nonumber \\
&\;\; -\frac{1}{2}\gamma ^a\epsilon [A_a,\bar \lambda \epsilon ^\prime ]+\frac{i}{2}\gamma ^a \epsilon [\bar \lambda \gamma _a\epsilon ^\prime ,\sigma ]+\frac{1}{2}\epsilon (\bar \lambda \epsilon ^\prime ) \nonumber \\
=&-2\bar \lambda (\epsilon \epsilon ^\prime )+i[\bar \lambda (\epsilon \epsilon ^\prime ),\sigma ]-[\gamma ^a\bar \lambda (\epsilon \epsilon ^\prime ),A_a]=0.
\end{align}
Thus it is indeed nilpotent.
The commutator between $\delta_{\epsilon}$ and $\delta_{\bar{\epsilon}}$
becomes a sum of gauge transformation, R-rotation and $SU(2)$ rotation, 
while the commutator for two barred or two unbarred parameters vanishes
(see appendix \ref{A.comm}).

We next consider the dimensional reduction of the YM term
for a gauge multiplet. By performing the above dimensional reduction
to (\ref{sym}), we obtain, 
\begin{align}
S^r_{YM}=&\Tr \Big[ \frac{1}{4}F_{ab}F^{ab}+\frac{1}{2}(\sigma +D)^2-\frac{1}{2}[A_a,\sigma ]^2+\frac{1}{2}\bar \lambda \gamma ^a[A_a,\lambda ]
 -\bar \lambda \lambda +\frac{i}{2}\bar \lambda [\sigma ,\lambda ]\Big].
\label{SYMaction}
\end{align}
This action can be written as a total superderivative:
\begin{align}
\bar \epsilon \epsilon S^r_{YM}=&\delta _{\bar \epsilon}\delta _{\epsilon}\Tr \! \left[ \frac{1}{2}\bar \lambda \lambda -2D\sigma \right].
\end{align}

We then consider a bifundamental matter multiplet 
$\{\phi, \psi, F\}$ coupled to two vector multiplets, 
$\{ A_a,\lambda ,\sigma ,D\}$ and 
$\{ B_a,\eta , \rho ,\tilde D\}$.
The matter action on $S^3$ is given by (\ref{smatt}).
Applying the same dimensional reduction to (\ref{smatt}), 
we obtain
\begin{align}
S^r_{matter}=& \Tr 
\Big[ \bar \phi \nabla (A_a,B_a)^2 \phi 
+\frac{3}{2}\bar{\psi} \psi -\bar{\psi} \gamma ^a\nabla (A_a,B_a)\psi 
+q(2-q)\bar{\phi} \phi -\frac{2q-1}{2}\bar{\psi} \psi 
\nonumber\\
&+i(2q-1)\bar{\phi} \nabla (\sigma ,\rho ) \phi 
+i\bar{\psi} \nabla (\sigma ,\rho )\psi 
+i\bar{\psi} \nabla (\lambda ,\eta )\phi 
-i\bar \phi \nabla (\bar \lambda ,\bar\eta)\psi 
\nonumber\\
&+i\bar \phi \nabla (D,\tilde D)\phi 
+\bar \phi \nabla (\sigma ,\rho )^2 \phi 
+\bar{F}F\Big].
\label{matter action}
\end{align}
Here, $\nabla (A,B)$ is defined in (\ref{nabla})
and $q$ is the anomalous dimension of the matter multiplet.
If the original matter action on $S^3$ has a superpotential, 
one can obtain a corresponding potential in the reduced 
model easily through the same procedure.
The action (\ref{matter action}) is 
invariant under the following supersymmetry transformations,
\begin{align}
\delta \phi =&\bar \epsilon \psi ,\nonumber\\
\delta \bar \phi =&\epsilon \bar \psi ,\nonumber\\
\delta \psi =& \gamma^a \epsilon \nabla(A_a,B_a)\phi 
+i\epsilon \nabla(\sigma,\rho) \phi 
-q\phi \epsilon +\bar \epsilon F ,\nonumber\\
\delta \bar{\psi} =&\gamma^a \epsilon \nabla(A_a,B_a) \bar\phi 
-i\nabla(\sigma,\rho)\bar{\phi} 
\bar\epsilon -q\bar \phi \bar \epsilon +\bar F\epsilon ,
\nonumber\\
\delta F=&(q-2)\psi \epsilon +\epsilon \gamma ^a\nabla(A_a,B_a)\psi 
-i\nabla (\sigma ,\rho ) \psi \epsilon 
-i(\epsilon \nabla(\lambda,\eta ) )\phi ,
\nonumber\\
\delta \bar F=&(q-2)\bar \psi \bar \epsilon 
+\bar \epsilon \gamma^a \nabla(A_a,B_a) \bar{\psi} 
+i\bar{\epsilon} \nabla (\sigma ,\rho )\bar{\psi}
-i(\bar{\epsilon} \nabla(\bar{\lambda},\bar{\eta} )) \bar{\phi}.
\label{matter susy}
\end{align}
One can also check the nilpotency on the matter fields. 
The matter action can also be written as a total superderivative:
\begin{align}
\bar \epsilon \epsilon S^r_{matter}=\delta _{\bar \epsilon}\delta _{\epsilon}\Tr [\bar \psi \psi -2i\bar \phi \nabla(\sigma,\rho) \phi +2(q-1)\bar \phi \phi ].
\end{align}

The adjoint matter is given as a special case of the bifundamental matter.
Namely, if we identify one gauge multiplet with the other, 
the action (\ref{matter action}) and the supersymmetry transformations 
(\ref{matter susy}) reduce to those for an adjoint matter multiplet.

The reduced model of a general ${\cal N}= 2$ quiver CS theory 
is constructed by combining (\ref{SCSaction}) and 
(\ref{matter action}) plus an appropriate
superpotential term \cite{Hanada:2009hd}. For example,
the ABJM theory contains two copies of supersymmetric CS theory 
of which gauge groups are $U(N_1)$ and $U(N_2)$ with opposite levels $k,-k$. 
The matter sector consists of 
four matter multiplets $\{ \phi _I,\psi _I, F_I\}$,
which are in the bifundamental representation for $I=1,3$
and in the anti-bifundamental for $I=2,4$. 
The anomalous dimension $q$ is $1/2$.
The action of the reduced model consists of
two CS terms (\ref{SCSaction})
and four copies of (\ref{matter action}) plus the superpotential term
which is obtained by the dimensional reduction of 
the quartic superpotential in the ABJM theory.

\subsection{Localization}
We then apply the localization to the reduced model of
a general ${\cal N}=2$ quiver CS theory.
We take a product of unitary groups $\bigotimes_{a}U(K_a)$ as the gauge 
group of the reduced model.

Since the reduced model preserves the nilpotent supersymmetry, 
one can perform the localization in the same manner as in the original 
theory on $S^3$;
we first add (\ref{SYMaction}) and (\ref{matter action}) 
to the action of the reduced model as 
$S^r \rightarrow S^r+tS^r_{YM}+t'S^r_{matter}$, 
where $t$ and $t'$ are parameters. 
Since (\ref{SYMaction}) and (\ref{matter action}) are 
exact under the supersymmetry,
the path integral does not depend on $t$ and $t'$.
Then, sending $t,t' \rightarrow \infty$
reduces the path integral to a sum of the 1-loop determinant
at each saddle point.

\subsubsection{Saddle points}
The localizing locus of the matrices in a gauge multiplet 
is determined by the vanishing condition of (\ref{SYMaction}),
\begin{align}
F_{ab}=0, \;\; \sigma +D=0, \;\;[A_a,\sigma ]=0.
\end{align} 
This is solved by 
\begin{align}
A_a=-2L_a, \;\; \sigma =\hat \sigma, \;\; D=-\hat \sigma ,
\label{gaugelocus}
\end{align}
where $L_a$ is a representation of $SU(2)$ generators 
obeying $[L_a,L_b]=i\varepsilon ^{abc}L_c $ and
$\hat{\sigma}$ satisfies
\begin{align}
[L_a,\hat \sigma ]=0.
\end{align}

$L_a$ is reducible in general and is decomposed to a direct sum
of irreducible representations in a suitable basis as
\begin{align}
L_a=
\begin{pmatrix}
\mathbf{1}_{M_{-\Lambda /2}}\otimes L_a^{[j_{-\Lambda /2}]}&&&&\\
&\ddots&&&\\
&& \mathbf{1}_{M_{s}}\otimes L_a^{[j_{s}]}&&\\
&&&\ddots&\\
&&&&\mathbf{1}_{M_{\Lambda /2}}\otimes  L_a^{[j_{\Lambda /2}]}
\end{pmatrix}.
\label{la}
\end{align}
Here $\Lambda$ is an even positive integer, 
$s=-\Lambda /2,-\Lambda /2+1,\cdots ,\Lambda /2$ label the diagonal blocks,
$L_a^{[j_s]}$ is the irreducible representation matrix of spin $j_s$, 
and $M_s$ is the multiplicity of each representation. 
The total matrix size of the gauge multiplet is given by 
$K=\sum_{s=-\Lambda /2}^{\Lambda /2}M_s(2j_s+1)$.
From Schur's lemma it follows that $\hat{\sigma}$ takes the following form 
in this basis,
\begin{equation}
\hat \sigma =
\begin{pmatrix}
\sigma _{-\Lambda /2}\otimes  \mathbf{1}_{2j_{-\Lambda /2}+1}&&&&\\
&\ddots&&&\\
&&\sigma _{s}\otimes  \mathbf{1}_{2j_{s}+1}&&\\
&&&\ddots&\\
&&&&\sigma _{\Lambda /2}\otimes \mathbf{1}_{2j_{\Lambda /2}+1}
\end{pmatrix},
\label{sigmahat}
\end{equation}
where $\sigma _{s}$ are $M_s\times M_s$ hermitian matrices.

Recall that in the original theory on $S^3$, all the gauge fields 
are zero at the saddle points up to gauge transformation.
When it is reduced to a point, however, the gauge equivalence class 
becomes smaller and hence the model becomes to possess many nontrivial 
saddle point configurations, (\ref{gaugelocus}).

It is easy to see that
at a saddle point, all the matrices in matter multiplets vanish.
They contribute to the partition function
only through the 1-loop determinant.

The path integral of the reduced model results in the 
integration over the moduli space of the saddle point configuration, 
which is given as the summation over the
representations of $SU(2)$ as well as the integration 
over $\hat{\sigma}$.
Thus the partition function takes the form
\begin{align}
Z=\sum_{\{ R_a \}}Z_{\{ R_a \}},
\label{zra}
\end{align}
where $R_a$ is a $K_a$ dimensional representation of $SU(2)$ as in 
(\ref{la}) for each gauge multiplet 
and the sum is taken over all possible representations for 
all the gauge multiplets.
$Z_{\{ R_a \}}$ is written as an integral over $\hat{\sigma}$ 
and the integrand is given by the product of the 1-loop determinants 
which come from all the multiplets. 

We make use of the residual gauge symmetry to 
diagonalize $\hat{\sigma}$. Then, the integration measure
obtains the Vandermonde determinant for each block labeled by $s$,
\begin{equation}
\int \dif{\hat \sigma}\to  \int \prod_{s,i}\dif{\sigma}_{si}
\prod_s \Delta(\sigma _s)^2, \;\;\; 
\Delta(\sigma _s)=\prod_{i> j}(\sigma _{si}-\sigma _{sj}),
\label{measure sigmahat}
\end{equation}
where $\sigma _{si}$'s are eigenvalues of $\sigma _s$.
In the following, we will work in this gauge.

\subsubsection{The gauge sector} 
We expand the matrices in a gauge multiplet 
around the saddle point given by (\ref{gaugelocus})
as $A_a = -2L_a +A_a'$, $\sigma=\hat{\sigma}+\sigma^\prime$ and so on,
and perform the 1-loop integral 
with respect to the fluctuations in the gauge multiplet.
For this purpose, we need to fix the residual gauge symmetry 
which leaves the background (\ref{gaugelocus}) invariant. 
From the original gauge symmetry
written as $A_a\to UA_aU^\dagger, \;\; U\in U(K)$, 
one can read off the transformation law for the 
fluctuation as
\begin{align}
A^\prime _a\to -2[U,L_a]U^\dagger +UA^\prime _aU^\dagger .
\end{align}
We adopt the standard Faddeev-Popov method and 
choose the following gauge-fixing condition for $A_a'$,
\begin{equation}
[L_a,A^{\prime \, a}]=0.
\end{equation}
Then the ghost and the gauge-fixing terms are given by
\begin{equation}
S^r_{ghost} =-\Tr \left( ib[L_a ,A^{\prime \, a}]+2\bar c[L_a,[L_a,c]]-
\bar c[L_a,[A^{\prime \, a},c]]\right).
\end{equation}
Note that the zero mode of the ghosts, that is, the mode satisfying
$[L_a, c]=0$ is absent in the above action.

We then perform the 1-loop integral around the saddle point keeping only 
the quadratic part of the fluctuation. 
The relevant part in the total action $S^r_{CS}+S^r_{YM}+S^r_{ghost}$ 
\footnote{We omit $t$ and $t'$ in front of the YM action and the 
matter action since they are irrelevant for the 
evaluation of the 1-loop determinant.}
for the gauge multiplet is given by 
\begin{align}
&\Tr \Big\{ (\varepsilon _{abc}A^{\prime \, c}+i[L_a,A^\prime _b]-i[L_b,A^\prime _a])^2+\frac{1}{2}(\sigma ^\prime +D^\prime )^2-\frac{1}{2}[\hat \sigma ,A^\prime _a]^2+[L_a,\sigma ^\prime ]^2\nonumber \\
&\;\;\;\;\;\;\;\; -\bar{\lambda}^\prime \gamma ^a[L_a,\lambda ^\prime  ]-\bar{\lambda}^\prime \lambda ^\prime  +\frac{i}{2}\bar{\lambda}^\prime  [\hat \sigma ,\lambda ^\prime  ] -2\bar c^\prime
[L_a,[L_a,c^\prime]]-ib^\prime[L_a,A^{\prime \, a}]\Big\} ,
\label{localizedaction}
\end{align}
where the fluctuations are represented by the primed matrices.
In (\ref{localizedaction}), the zero mode of $\sigma^{\prime}$
is not included. Since this mode corresponds to the direction of the 
moduli of the saddle point, it is treated in the moduli 
integral in (\ref{measure sigmahat}).

The integration over $\sigma ^\prime$, $D^\prime$,
$c^\prime,\bar c^\prime$ and $b^\prime$ yields  
\begin{align}
\det {}^{\prime} ([L_a,[L_a,\cdot ]])^{1/2} \delta ([L_a,A^{\prime \, a}]),
\label{cbsigma}
\end{align} 
up to an overall constant, where $\det^{\prime}$ means that 
the zero mode is removed in taking the determinant.

In order to calculate further, let us decompose the matrices to smaller 
blocks which are defined by the structure of $L_a$ in (\ref{la}).
We label each block by a pair $(s,t)$, where
$s,t$ run from $-\Lambda/2$ to $\Lambda/2$,
and denote the $(s,t)$ block of $A_a^\prime$ by 
$A_a^{\prime \, (s,t)}$.
Note that $A_a^{\prime \, (s,t)}$ is a
$(2j_s+1)M_s \times (2j_t+1)M_t$ rectangular matrix.
Each block component can be expanded in terms of the vector 
fuzzy spherical harmonics 
defined in appendix \ref{A.harmonics} as
\begin{equation}
A_a^{\prime \, (s,t)}=\sum_{\rho =-1}^{1}\sum_{\tilde Q= |j_s-j_t|}^{j_s+j_t}\sum_{m=-Q}^{Q}a_{Jm\rho}^{(s,t)}\otimes \hat Y^\rho _{Jm(j_sj_t)a},
\label{harmonic A}
\end{equation}
where $Q=J+\delta _{\rho ,1}, \tilde Q=J+\delta _{\rho ,-1}$ and 
the sum over $\tilde{Q}$ is taken for $J\geq 0$. 
$a_{Jm\rho}^{(s,t)}$ is a matrix with size $M_s \times M_t$.

The delta function in (\ref{cbsigma}) constrains
the $\rho =0$ component, $a^{(s,t)}_{Jm0}$, 
in the above expansion to vanish. 
In fact, using \eqref{LYvec}, we find that 
\begin{align}
\delta ([L_a,A^{\prime \, a}])=
\prod_{s,t}\prod_{\substack{J\geq |j_s-j_t| \\ J\neq 0} }^{j_s+j_t}
\prod_{m=-J}^{J}\{ \sqrt{J(J+1)}\} ^{-1}\delta (a^{(s,t)}_{Jm0}).
\end{align}
We can therefore integrate out $a^{(s,t)}_{Jm0}$ trivially 
leaving the factor,
\begin{equation}
\prod_{s,t}\prod_{\substack{J\geq |j_s-j_t| \\ J\neq 0}}^{j_s+j_t}
\prod_{m=-J}^{J}
\{ \sqrt{J(J+1)}\} ^{-1}=\{ \det {}^\prime ([L_a,[L_a,\cdot ]])\} 
^{-\frac{1}{2}}.
\end{equation}
This cancels the other factor in (\ref{cbsigma}).

We then perform the integral over $A_a^\prime$ with 
$\rho =\pm 1$ in the expansion (\ref{harmonic A}).
By substituting (\ref{harmonic A}) to (\ref{localizedaction}), 
the relevant part of the action becomes
\begin{align}
&\Tr \Big\{ (\varepsilon _{abc}A^{\prime \, c}+i[L_a,A^\prime _b]-i[L_b,A^\prime _a])^2-\frac{1}{2}[\hat \sigma ,A^\prime _a]^2\Big\} \nonumber \\
=&\sum_{s,t}\sum_{i,j}\sum_{\rho =\pm 1}\sum_{J,m}
\left| ( a^{(s,t)}_{Jm\rho })_{ij}\right| ^2\Big[ 2(J+1)^2+\frac{1}{2}(\sigma _{si}-\sigma _{tj})^2\Big] ,
\end{align}
where $i=1,2,\cdots, M_s$ and $j=1,2,\cdots, M_t$.
We have used the formulae (\ref{LY+Yvec}), (\ref{YvecHC}) 
and (\ref{trYvec}).
Then the integration results in the factor,
\begin{equation}
\prod_{s,t}\prod_{i,j}\prod_{\rho =\pm 1}\prod_{J,m}\left[ 2(J+1)^2+\frac{1}{2}(\sigma _{si}-\sigma _{tj})^2\right] ^{-1/2}.
\end{equation}
The exponent $-1/2$ comes from the fact that $a^{(s,t)}_{Jm\rho}$ 
satisfy a kind of the reality condition, 
\begin{equation}
a^{(s,t)}_{Jm\rho}=(-1)^{m-(j_s-j_t)+1}a^{(t,s)\, \dagger}_{J\, -m\rho},
\end{equation}
which follows from (\ref{YvecHC}) and the Hermiticity of $A_a^\prime$.

To perform the integration over $\lambda ^\prime $ and 
$\bar{\lambda} ^\prime $, we also expand them in terms 
of the spinor fuzzy spherical harmonics defined in 
(\ref{spinor harmonics}) as
\begin{align}
\lambda ^{\prime \, {(s,t)\alpha}}=&\sum_{\kappa=\pm 1}
\sum_{\tilde U= |j_s-j_t|}^{j_s+j_t}\sum_{m=-U}^{U}\lambda ^{(s,t)}_{Jm\kappa}
\otimes \hat Y^\kappa _{Jm(j_sj_t)\alpha},\\
\bar{\lambda}^{\prime \, (s,t)}_\alpha=&\sum_{\kappa=\pm 1}
\sum_{\tilde U= |j_s-j_t|}^{j_s+j_t}\sum_{m=-U}^{U}\bar \lambda ^{(s,t)}_{Jm\kappa}
\otimes \left( \hat Y^\kappa _{Jm(j_sj_t)\alpha}\right) ^\dagger ,
\end{align}
where $\alpha$ denotes the spinor index, $U=J+\frac{1}{2}\delta _{\kappa ,1}, \tilde U=J+\frac{1}{2}\delta _{\kappa ,-1}$ and the summation over $\tilde{U}$
is taken for $J\geq 0$.
Plugging these expansions into the action, we obtain 
\begin{align}
&\Tr \left\{ -\bar{\lambda}^\prime \gamma ^a[L_a,\lambda ^\prime  ]-\bar{\lambda}^\prime \lambda ^\prime  +\frac{i}{2}\bar{\lambda}^\prime  [\hat \sigma ,\lambda ^\prime  ]\right\}
\nonumber\\
=&\sum_{s,t}\sum_{i,j}\sum_{\kappa =\pm 1}\sum_{J,m}
(\bar \lambda ^{(s,t)}_{Jm\kappa})_{ij}
(\lambda ^{(t,s)}_{Jm\kappa})_{ji}
\left[ -\frac{1}{4}-\kappa \left(J+\frac{3}{4}\right)
-\frac{i}{2}(\sigma _{si}-\sigma _{tj})\right] ,
\end{align}
where we have used the formulae \eqref{LY+Yspn} and (\ref{trYspn}).
Therefore, the integration with respect to $\lambda^\prime$ 
and $\bar\lambda^\prime$ gives the factor,
\begin{equation}
\prod_{s,t}\prod_{i,j}\prod_{\kappa =\pm 1}\prod_{J,m}\left[ \frac{1}{4}+\kappa \left(J+\frac{3}{4}\right)+\frac{i}{2}(\sigma _{si}-\sigma _{tj})\right].
\end{equation}

Combining the above results with the Vandermonde determinant for 
$\hat{\sigma}$, we obtain the 1-loop determinant 
from the gauge multiplet,
\begin{align}
\prod_{s}\Delta(\sigma_s)e^{-\sum_{s,i}\frac{2j_s+1}{g^2}\sigma_{si}^2}
&\frac{\prod_{s\neq t}\prod_{i,j}
\prod_{J=|j_s-j_t|}^{j_s+j_t} 
(-J-1+i(\sigma_{si}-\sigma_{tj})/2)^{2J+2}}
{\prod_{s < t}\prod_{i,j}
\prod_{J=|j_s-j_t|}^{j_s+j_t} 
((J+1)^2+(\sigma_{si}-\sigma_{tj})^2/4)^{2J+3}}
\nonumber\\
\times & \frac{\prod_{s\neq t}\prod_{i,j}
\prod_{J=|j_s-j_t|-1/2}^{j_s+j_t-1/2} 
(J+1/2+i(\sigma_{si}-\sigma_{tj})/2)^{2J+1}}
{\prod_{s < t}\prod_{i,j}
\prod_{J=|j_s-j_t|-1}^{j_s+j_t-1} 
((J+1)^2+(\sigma_{si}-\sigma_{tj})^2/4)^{2J+1}
}.
\end{align}
Because of the cancellation between bosons and fermions,
this is simplified to 
\begin{align}
e^{-\sum_{s,i}\frac{2j_s+1}{g^2}\sigma_{si}^2}
\prod_{s}\prod_{i<j} 
\frac{(\sigma_{si}-\sigma_{sj})^2}
{(2j_s+1)^2+(\sigma_{si}-\sigma_{sj})^2/4}
\prod_{s<t}\prod_{i,j}
\frac{(j_s-j_t)^2+(\sigma_{si}-\sigma_{tj})^2/4}
{(j_s+j_t+1)^2+(\sigma_{si}-\sigma_{tj})^2/4}.
\label{gauge 1-loop}
\end{align}

\subsubsection{The matter sector}
We first consider the matter multiplet in the bifundamental representation
(\ref{matter action}). 
After integrating out $F$ and $\bar{F}$ trivially, 
the relevant part of the matter action is given by\footnote{We will omit the
primes for the fluctuations of the matters.}
\begin{align}
&\Tr \Big[ \bar \phi \{ 4 \nabla (L_a,\tilde L_a)^2+1+\left( \nabla (\hat \sigma ,\hat \rho ) -i(1-q)\right) ^2\} \phi \nonumber \\
&\;\;\;\;\;\;\; +\bar \psi \{ 2\gamma ^a\nabla (L_a,\tilde L_a) +i\nabla (\hat \sigma ,\hat \rho )+(2-q)\} \psi \Big].
\end{align}
Here, $\tilde{L}_a$ and $\hat{\rho}$ are 
the saddle point configurations for the second 
gauge multiplet and are the counterparts of 
$L_a$ and $\hat{\sigma}$, respectively.
They take a similar form to (\ref{la}) and (\ref{sigmahat}).
We define $t$, $k_t$, $\tilde{\Lambda}$ and $\tilde{M}_t$ for 
$\tilde{L}_a$ and $\hat{\rho}$
as the counterparts of $s$, $j_s$, $\Lambda$ 
and $M_s$ in (\ref{la}) and (\ref{sigmahat}), respectively.

We decompose the bifundamental (rectangular) matrices to 
the block components and denote them by 
$\phi ^{(s,t)}$ and $\psi ^{(s,t)}$
corresponding to the zz in $L_a$ and the $t$-th block  
in $\tilde L_a$.
They are $(2j_s+1)M_s \times (2k_t+1)\tilde{M}_t$ matrices and
can be expanded in terms of the fuzzy spherical harmonics as
\begin{align}
\phi ^{(s,t)}&=\sum_{J=|j_s-k_t|}^{j_s+k_t}\sum_{m=-J}^{J}
\phi ^{(s,t)}_{J m}\otimes \hat Y_{J m(j_sk_t)},
\nonumber\\
\psi ^{(s,t)}_{\alpha}&=\sum_{\kappa=\pm 1} 
\sum_{\tilde{U}=|j_s-k_t|}^{j_s+k_t}\sum_{ m=-U}^{U}
\psi ^{(s,t)}_{J m\kappa}\otimes \hat Y_{J m(j_sk_t)\alpha}^{\kappa},
\label{expansion for phi}
\end{align}
where $\phi ^{(s,t)}_{J\tilde m}$ and $\psi ^{(s,t)}_{J m\kappa}$ are 
the $M_s\times \tilde M_t$ matrices.
In this basis, $\nabla (L_a,\tilde L_a)$ can be rewritten as
\begin{equation}
\nabla (L_a,\tilde L_a)\phi ^{(s,t)}=\sum_{J=|j_s-k_t|}^{j_s+k_t}\sum_{\tilde m=-J}^{J}\phi ^{(s,t)}_{J\tilde m}\otimes L_a\circ \hat Y_{J\tilde m(j_sk_t)},
\end{equation}
where $L_a\circ$ is defined in \eqref{LcM}. 
By substituting (\ref{expansion for phi}), 
the quadratic part of the matter action becomes
\begin{align}
&\sum_{s,t}\sum_{i,\alpha} \sum_{J=|j_s-k_t|}^{j_s+k_t}
\sum_{m=-J}^{J} (\bar{\phi}^{(t,s)}_{J m})_{\alpha i} 
(\phi ^{(s,t)}_{J m})_{i\alpha}
\left[ (2J+1)^2+\left( \sigma _{si}-\rho _{t\alpha}-i(1-q)\right) ^2\right]
\nonumber\\ 
+ &\sum_{s,t}\sum_{i,\alpha}\sum_{\kappa=\pm 1}
\sum_{\tilde{U}=|j_s-k_t|}^{j_s+k_t}
\sum_{m=-U}^{U}(\bar{\psi}^{(t,s)}_{Jm\kappa})_{\alpha i}
(\psi ^{(s,t)}_{Jm})_{i \alpha}
\left[ 2\kappa \left(J+\frac{3}{4}\right)-\left(q-\frac{1}{2}\right)
+i\left( \sigma _{si}-\rho _{t\alpha}\right) \right],
\end{align}
where $i=1,2,\cdots ,M_s$ and 
$\alpha =1,2,\cdots, \tilde M_t$\footnote{
It will not cause any confusion 
to use $\alpha$ both for the spinor index and 
for the index labeling the diagonal components
of $\rho_{t}$.}.
Then, we find after a simple calculation that 
the 1-loop determinant for the bifundamental matter multiplet is given by
\begin{align}
\prod_{s,t}\prod_{i,\alpha}\prod_{J=|j_s-k_t|}^{j_s+k_t}
\frac{2J+2-q+i(\sigma _{si}-\rho _{t\alpha})}
{2J+q-i(\sigma _{si}-\rho _{t\alpha})}.
\end{align}

The 1-loop determinant from the adjoint matter is easily obtained 
by identifying one gauge multiplet with the other in the above calculation. 
The result is given by
\begin{equation}
\prod_{s,t}\prod_{i>j}\prod_{J=|j_s-j_t|}^{j_s+j_t}\frac{(2J+2-q)^2+(\sigma _{si}-\sigma _{tj})^2}{(2J+q)^2+(\sigma _{si}-\sigma _{tj})^2}.
\label{adparti}
\end{equation}

In the reduced model of the ABJM theory, $q=1/2$ and there are 
two bifundamental and two anti-bifundamental matters.
Hence the 1-loop determinant from the matter sector is given by
\begin{align}
\prod_{s,t}\prod_{i,\alpha}\prod_{J=|j_s-k_t|}^{j_s+k_t}\left( \frac{(2J+\frac{3}{2})^2+(\sigma _{si}-\rho _{t\alpha})^2}{(2J+\frac{1}{2})^2+(\sigma _{si}-\rho _{t\alpha})^2}\right) ^2.
\label{matterparti}
\end{align}

\subsubsection{Wilson loop}
The Wilson loop operator in the reduced models of theories on $S^3$ 
was constructed in \cite{Ishii:2007sy,Ishiki:2011ct}.
It is given as a naive dimensional reduction of the Wilson loop in 
the theory on $S^3$ (\ref{2.6}),
\begin{align}
\hat{W}(\mathcal{C})=\frac{1}{K}\Tr \left[ P\exp 
\left( i\oint_{\mathcal{C}} \dif{s}(\dot x^\mu (s)e_\mu ^{\; a}(x)
A_a-i|\dot x(s)|\sigma ) \right) \right].
\end{align}
In the case that the contour is a great circle on $S^3$,
this operator is BPS in the reduced model, so that
we can calculate the correlation function of this operator 
by the localization technique. 
In this case, substituting (\ref{2.7}) simplifies the operator as
\begin{align}
\hat{W}(\mathcal{C})=\frac{1}{K}\Tr \left[ 
e^{ 2\pi i(A_3-i\sigma )} \right].
\end{align}
Then applying the localization, we obtain
\begin{align}
\langle \hat{W}(\mathcal{C})\rangle 
=&\frac{1}{K} \langle \Tr 
e^{-4\pi iL_3+2\pi \hat \sigma}\rangle \\
=&\frac{1}{K}\sum_{s=-\Lambda /2}^{\Lambda /2} (2j_s+1)\sum_{i=1}^{M_s}\langle e^{2\pi \sigma _{si}}\rangle ,
\label{MMWilson}
\end{align}
where to obtain the second line we have used 
the fact that each diagonal component 
of $L_3^{[j_s]}$ takes a value in either integer or half-integer.
$\langle \cdots \rangle $ stands for an average with respect to the 
eigenvalue integral for the reduced model.

\section{$\mathcal{N}=2$ quiver CS theory on $S^3$ from reduced model}
In this section, from the reduced model we realize a quiver CS theory on $S^3$ 
with gauge group $\bigotimes_aU(N_a)$ in the 't Hooft limit in which 
\begin{align}
N_a\rightarrow \infty \quad \text{with} \quad 
\frac{N_a}{N_b} \quad \text{and} \quad \frac{N_a}{k_a} \quad \text{fixed}
\label{thooft lim in cs}
\end{align}
for any $a$ and $b$.
We assume for simplicity the gauge group to be $U(N_1)\times U(N_2)$ 
and mainly consider the case of the ABJM theory, 
which contains two bifundamental multiplets $(N_1,\bar{N}_2)$ 
and two anti-bifundamental multiplets $(\bar{N}_1, {N}_2)$.
However, it will be obvious that 
our argument is applicable to more general 
$\mathcal{N}=2$ quiver CS theories.
To realize the theory on $S^3$, we take the reduced model with gauge group $U(K_1)\times U(K_2)$
where $K_i$'s ($i=1,2$) are much larger than $N_i$'s such that the Kaluza-Klein (KK) modes on $S^3$
are embedded in matrices in the reduced model.
We denote by $g_1$ and $g_2$ the coupling constants for 
the two CS terms (\ref{SCSaction}) in the reduced model.

\subsection{$S^3$ from matrices}

Recall that the partition function in the reduced model is given 
by a sum of 1-loop contributions around saddle points, 
each of which is specified by a representation of $SU(2)$, (\ref{zra}). 
In order to obtain the theory on $S^3$, 
we extract the following representations from the sum in (\ref{zra}),
\bea
A_a=-2 \bigoplus_s L^{[j_s]}_a \otimes 1_{N_1}, \quad 
B_a=-2 \bigoplus_s L^{[j_s]}_a \otimes 1_{N_2},
\label{rep for sp}
\eea
which corresponds to the case of $M_s=N_1$ for $A_a$ and 
$\tilde{M}_s=N_2$ for $B_a$ in (\ref{gaugelocus})
and $K_i=N_i\sum_s(2j_s+1)$.
We take $j_s$ as 
\bea
2j_s+1 = n +s  \ \ \ \text{for\ } -\frac{\Lambda}{2} \leq s \leq \frac{\Lambda}{2}\
\label{rep for sphere},
\eea
where $n$ is a positive integer.
We then take the limits in which
\begin{align}
&n \rightarrow \infty,\quad \frac{g_1^2}{n}\rightarrow 0,\quad \frac{g_2^2}{n}\rightarrow 0, \n
&\Lambda  \rightarrow \infty,\quad n-\Lambda \rightarrow \infty,  \n
&N_{1} \rightarrow {\infty},\quad N_{2} \rightarrow {\infty}
\label{planar limit}
\end{align}
with the following combinations fixed
\begin{align}
t_1 \equiv \frac{N_1 g^2_1}{n}, \quad t_2 \equiv \frac{N_2 g^2_2}{n}, \quad \frac{N_1}{N_2}.
\label{planar limit fix}
\end{align}

Here we explain the reason why the above representation and the limit create $S^3$ (see \cite{Ishii:2008ib} for more detail). 
First, we consider the original theory on $S^3$.
Since $S^3$ is viewed as an $S^1$-bundle over $S^2$,
we can perform the KK reduction along the fiber direction.
Then we obtain a theory on $S^2$ involving infinite KK modes.
Reflecting the nontrivial fibration of $S^1$,
the KK mode with KK momentum $\tilde{m}$ on $S^2$ 
can be regarded as a field in a monopole background,
where the monopole is sitting at the center of $S^2$ in $R^3$ with monopole charge $\tilde{m}$.
As the angular momentum of the field on $S^2$ in the presence of a monopole is bounded below
by its charge $J\geq |\tilde{m}|$, that of the KK mode is also bounded.
The same situation can be observed in the mode expansion of a rectangular 
matrix (\ref{expansion for phi})
if one identifies the angular momentum $J$ and the monopole charge $\tilde{m}$ on $S^2$ 
with $J$ and $j_s-j_t$ in (\ref{expansion for phi}), respectively.
The only difference is the existence of the upper bound of the angular momentum
$j_s+j_t$, which can be removed 
by putting $2j_s+1=n+s$ and taking the $n\rightarrow \infty$ limit
so that $j_s+j_t\rightarrow \infty$ and $j_s-j_t=\frac{s-t}{2}=\tilde{m}$.
Indeed the rectangular block of the fluctuation is a regularization of a field on $S^2$
in a monopole background.
Thus, expanding the reduced model around the appropriate representation \eqref{rep for sp} 
such that the full KK modes ($-\infty\leq \tilde{m}\leq \infty$) on $S^2$ are reproduced,
we can obtain the original theory on $S^3$.
Now the physical interpretation of $n$ and $\Lambda$ is clear;
$n$ plays a role of UV momentum cutoff on $S^2$ while $\Lambda$ plays a role of UV cutoff on $S^1$.
Therefore $\{ L_a^{[j_s]} \}$ in \eqref{rep for sp} creates $S^3$ while 
the multiplicities $N_1$ and $N_2$ reproduce
the original gauge group $U(N_1)\times U(N_2)$.

In the following calculation of the partition function or 
the Wilson loop, 
we first take the $n\rightarrow \infty$ limit
shown in \eqref{planar limit}
with $n/g_1^2$ and $n/g_2^2$ fixed and later we take the other limits.
This is possible since the $n\rightarrow \infty$ limit does not lead to
any divergence.
In the $n\rightarrow \infty$ limit,
we can replace the coefficients of the Gaussian terms 
in (\ref{gauge 1-loop}) with 
$\frac{n}{g_i^2}=\frac{N_i}{t_i}$.
Then the contribution of the representation \eqref{rep for sp} from the summation 
in (\ref{zra}) is given by
\bea
\int \prod_{s=-\halfL}^{\halfL}\left(
\prod_{i=1}^{N_1} d \s_{si}
\prod_{\alpha=1}^{N_2} d\rho_{s\alpha}\right)\ 
 \mathcal{M}_{gauge}\ \mathcal{M}_{matter} \
\exp\left(-\frac{N_1}{t_1}\sum_{s,i} \sigma_{si}^2
+\frac{N_2}{t_2}\sum_{s,\alpha}\rho_{s\alpha}^2\right),
\label{sphere action}
\eea
where $t_1$ and $t_2$ are defined in (\ref{planar limit fix})
and $\mathcal{M}_{gauge}$ and $\mathcal{M}_{matter}$ are 1-loop determinants 
for the gauge and the matter sector, respectively, in the case of 
(\ref{rep for sp}).

More concretely, $\mathcal{M}_{gauge}$ is given by
\begin{align}
\mathcal{M}_{gauge}
&= \prod_s \prod_{i<j}(\s_{si}-\s_{sj})^2 \prod_{s<t}\prod_{i,j}
\left[1+\frac{(\s_{si}-\s_{tj})^2}{(s-t)^2}\right]
 \non\\
 & \quad \times  
\prod_s \prod_{\al<\beta}(\rho_{s\al}-\rho_{s\beta})^2\prod_{s<t}\prod_{\al,\beta}
 \left[1+ \frac{(\rho_{s\al} -\rho_{t\beta})^2}{(s-t)^2}\right], 
 \label{measure from vector}
\end{align}
where we have dropped the denominator in (\ref{gauge 1-loop}) 
because it becomes independent of $\s_{si}$ or $\rho_{s\alpha}$ 
in the limit, $n\rightarrow \infty$.  
In addition we have dropped the irrelevant constant 
factor $\prod_{s<t}(s-t)^{2N_1^2+2N_2^2}$.

$\mathcal{M}_{matter}$ depends on the matter content.
For a matter multiplet in the bifundamental representation,  
it is given by 
\bea
\mathcal{M}_{matter} \biggr|_{bifund.}=\prod_{s,t}  \prod_{i,\al} \prod_{J=|s-t|/{2}}^{\infty} \frac{2J+2-q+i(\s_{si}-\rho_{t\al})}{2J+q-i(\s_{si}-\rho_{t\al})}, \label{measure of a single matter}
\eea
and for the ABJM theory, it is given by
 \bea
\mathcal{M}_{matter}\biggr|_{ABJM}
=\prod_{s,t} \prod_{i,\al} \prod_{J=|s-t|/{2}}^{\infty} 
\biggl(
\frac{(2J+\frac{3}{2})^2+(\s_{si}-\rho_{t\al})^2}{(2J+\frac{1}{2})^2+(\s_{si}-\rho_{t\al})^2}
\biggr)^2.
\label{measure from matter}
\eea

\subsection{Perturbative proof of large-$N$ equivalence}
\subsubsection{Feynman rule for reduced matrix model}
We consider the perturbation theory of (\ref{sphere action}) 
with respect to the 't Hooft couplings, $t_1$ and $t_2$,
in the limit \eqref{planar limit}. 
Here $\mathcal{M}_{gauge}$ and $\mathcal{M}_{matter}$ are regarded as 
interactions. 
We will prove the equivalence between 
the reduced model \eqref{sphere action} and the original theory on $S^3$ 
by showing one to one correspondence of Feynman diagrams between these theories.

To read off the Feynman rules, 
it is convenient to rewrite \eqref{sphere action} 
in a manifestly $U(N_1)\times U(N_2)$ invariant form, 
which is given by a multi-matrix model consisting of matrices $\s_s$ and $\rho_s$ 
with double trace interactions.

In \eqref{measure from vector},
the factors $\prod_s \prod_{i<j} (\s_{si}-\s_{sj})^2$ 
and $\prod_s \prod_{\al<\beta}(\rho_{s\al}-\rho_{s\beta})^2$ correspond to 
the Vandermonde determinants for matrices $\sigma_s$ and $\rho_s$, and
the remaining factor of $\sigma_{si}$ can be written as
\begin{align}
\prod_{s<t}\prod_{i,j}
\biggl[1+ \frac{(\s_{si} -\s_{tj})^2}{(s-t)^2}\biggr]
&=\exp\left[
\frac{1}{2}\sum_{s\neq t}\sum_{i,j}\ln \left\{1+\frac{(\sigma_{si}-\sigma_{tj})^2}{(s-t)^2}\right\}
\right] \n
&=\exp
\biggl[
-\sum_{s\neq t} \underset{a+b\in 2 \IN}{\sum_{a,b\in \mathbb{Z}_{\geq 0}}}
\frac{K_{ab}}{(s-t)^{a+b}}
\tr  \s_s^a\ \tr \s_t^b
\biggr],
\label{sigma int from vector measure}
\end{align}
where $\mathbb{Z}_{\geq 0}$ is the set of non-negative integers 
and $2\mathbb{N}$ is the set of even positive integers.
$K_{ab}$ is a numerical factor given by
\bea
K_{ab}\equiv \frac{(-1)^{\frac{a+b}{2}}}{a+b}
\begin{pmatrix}
a+b\\
a \\
\end{pmatrix}
(-1)^{a}.
\label{Kab}
\eea
The factor consisting of $\rho_{s\al}$ in \eqref{measure from vector} is obtained 
by just replacing $\sigma\rightarrow \rho$ in \eqref{sigma int from vector measure}.

For a matter multiplet in the bifundamental representation, 
if one naively applies the same calculation to the factor \eqref{measure of a single matter}, one obtains
\begin{align}
\exp\biggl[-
\sum_{s,t}\sum_{J=\frac{|s-t|}{2}}^{\infty}\sum_{n=1}^{\infty}\frac{(-1)^n}{n}
\biggl\{
\left(\frac{i}{2J+2-q}\right)^n-\left(\frac{-i}{2J+q}\right)^n
\biggr\}
\sum_{r=0}^{n}
\begin{pmatrix}
n\\
r \\
\end{pmatrix}
(-1)^{n-r}
\tr \s_s^r \ \tr \rho_t^{n-r}
\biggr]. \label{bifund int in reduced}
\end{align}
We find that for $n=1$, the coefficients of $\tr \s_s$ and 
$\tr \rho_s$, are divergent
since  they have the form $\sum_J \frac{1}{J}$, 
and therefore we can not perform the perturbative expansion. 
However, there is no such a divergence in non-chiral theories such as the ABJM theory,
which we consider below.
Note that if we restrict $\s_s $ and $ \rho_s $ to 
traceless matrices for each $s$, $\tr \s_s = \tr \rho_s =0$,
we do not  have the divergence. 
In this case the following argument can be applied 
and the reduced model properly realizes the perturbative expansion in
${\cal N}=2$ quiver CS theory with a bifundamental matter of 
$SU(N_1) \times SU(N_2)$ gauge group .

For the reduced model of the ABJM theory, 
$\mathcal{M}_{matter}$ can be written as
\begin{align}
\exp
\Biggl[4
\sum_{s,t}\underset{a+b\in 2\IN}{\sum_{a,b\in \mathbb{Z}_{\geq 0}}}
\frac{K_{ab}}{2^{a+b}}
\biggl\{\zeta\left(a+b, \frac{1}{4}+\frac{|s-t|}{2}\right)
-\zeta\left(a+b,\frac{3}{4}+\frac{|s-t|}{2}\right)
\biggr\}
\tr \s_s^a \tr \rho_t^b
\Biggr],
\end{align}
where $\zeta(z,\ q)$ is the generalized zeta function
\bea
\zeta(z,\ q)= \sum_{n=0}^{\infty}\frac{1}{(q+n)^z}.\label{zeta}
\eea 

In summary, the reduced model of the ABJM theory is given by
\bea
\int \Biggl( \prod_{s=-\frac{\Lambda}{2}}^{\frac{\Lambda}{2}}d\s_s d \rho_s \Biggr) 
\exp\left( -\frac{N_1}{t_1}\sum_s  \tr \s_s^2 +\frac{N_2}{t_2}\sum_s\tr \rho_s^2 
+U^r_{gauge}+U_{matter}^r \right), \label{reduced model action}
\eea
where $U_{gauge}^r$ and $U_{matter}^r$ are the double-trace interactions:
\begin{align}
U_{gauge}^r&= -\sum_{s\neq t} \underset{a+b\in 2 \IN}{\sum_{a,b\in \mathbb{Z}_{\geq 0}}}
\frac{K_{ab}}{(s-t)^{a+b}}
(\tr  \s_s^a\ \tr \s_t^b+\tr  \rho_s^a\ \tr \rho_t^b ), \\
U_{matter}^r&=
\sum_{s,t}\underset{a+b\in 2\IN}{\sum_{a,b\in \mathbb{Z}_{\geq 0}}}
\frac{4K_{ab}}{2^{a+b}}
\left\{\zeta\left(a+b, \frac{1}{4}+\frac{|s-t|}{2}\right)
-\zeta\left(a+b,\frac{3}{4}+\frac{|s-t|}{2}\right)
\right\}
\tr \s_s^a \tr \rho_t^b. \label{reduced model int}  
\end{align}

\begin{figure}[]
\begin{center}
\subfigure[ $\s$-propagator]{\includegraphics*[width=.20\linewidth]{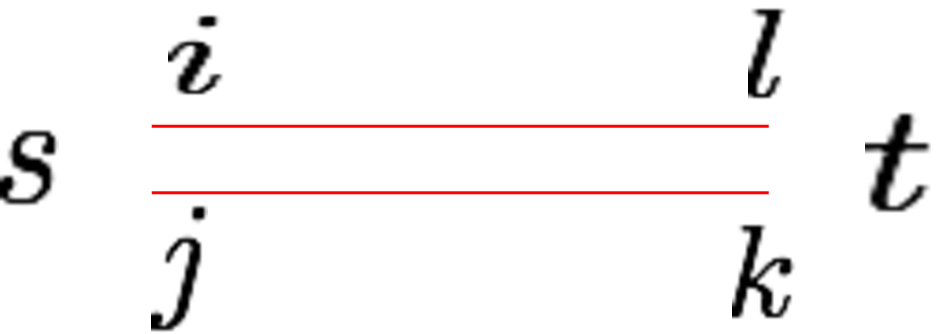}
\label{fig:p_s}}
\subfigure[ $\rho$-propagator]{\includegraphics*[width=.20\linewidth]{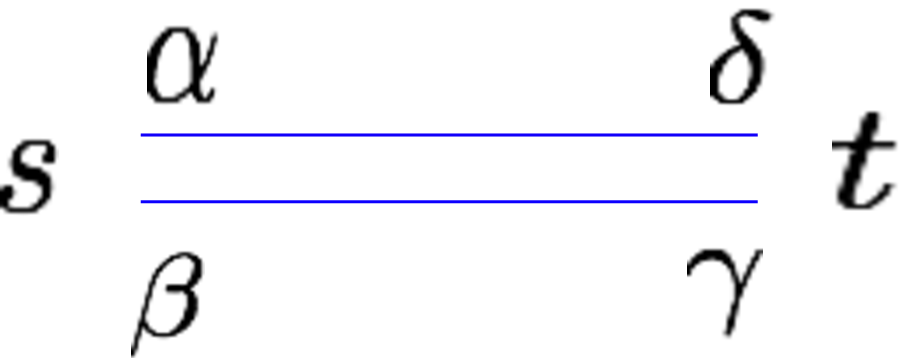}
\label{fig:p_r}}
\subfigure[ $\tr  \s_s^6\ \tr \s_t^2$]{\includegraphics*[width=.20\linewidth]{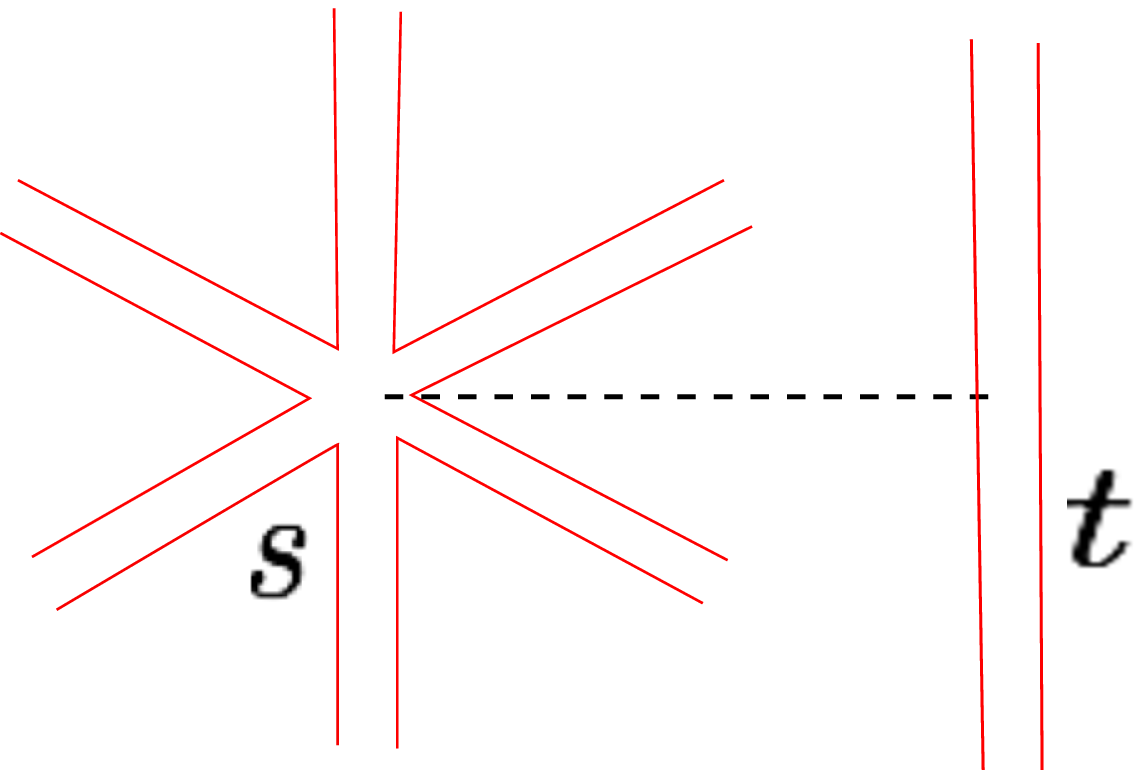}
\label{fig:s6s2}} \\
\subfigure[ $\tr  \rho_s^4\ \tr \rho_t^4$]{\includegraphics*[width=.25\linewidth]
{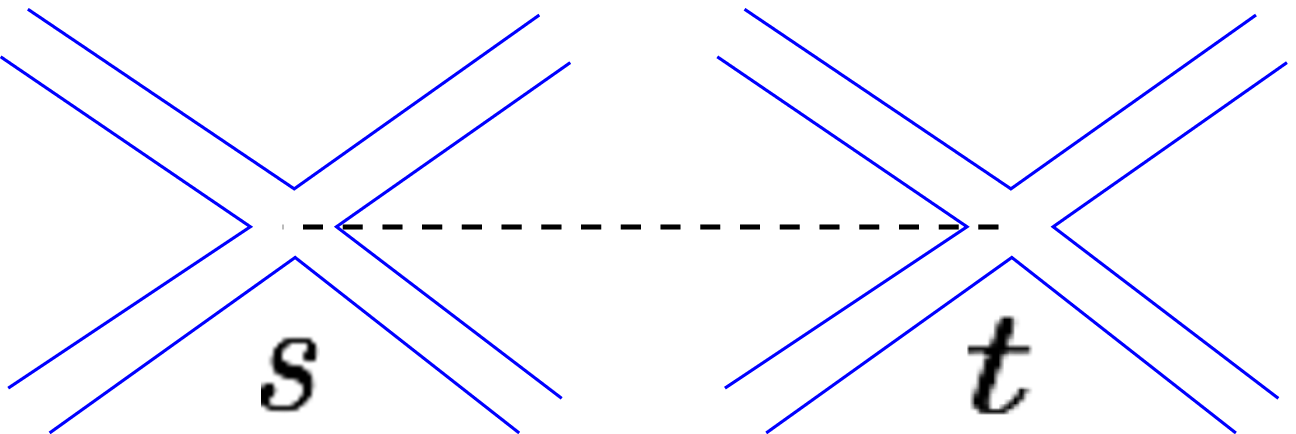}
\label{fig:r4r4}}
\subfigure[ $\tr  \s_s^6\ \tr \rho_t^4$]{\includegraphics*[width=.25\linewidth]{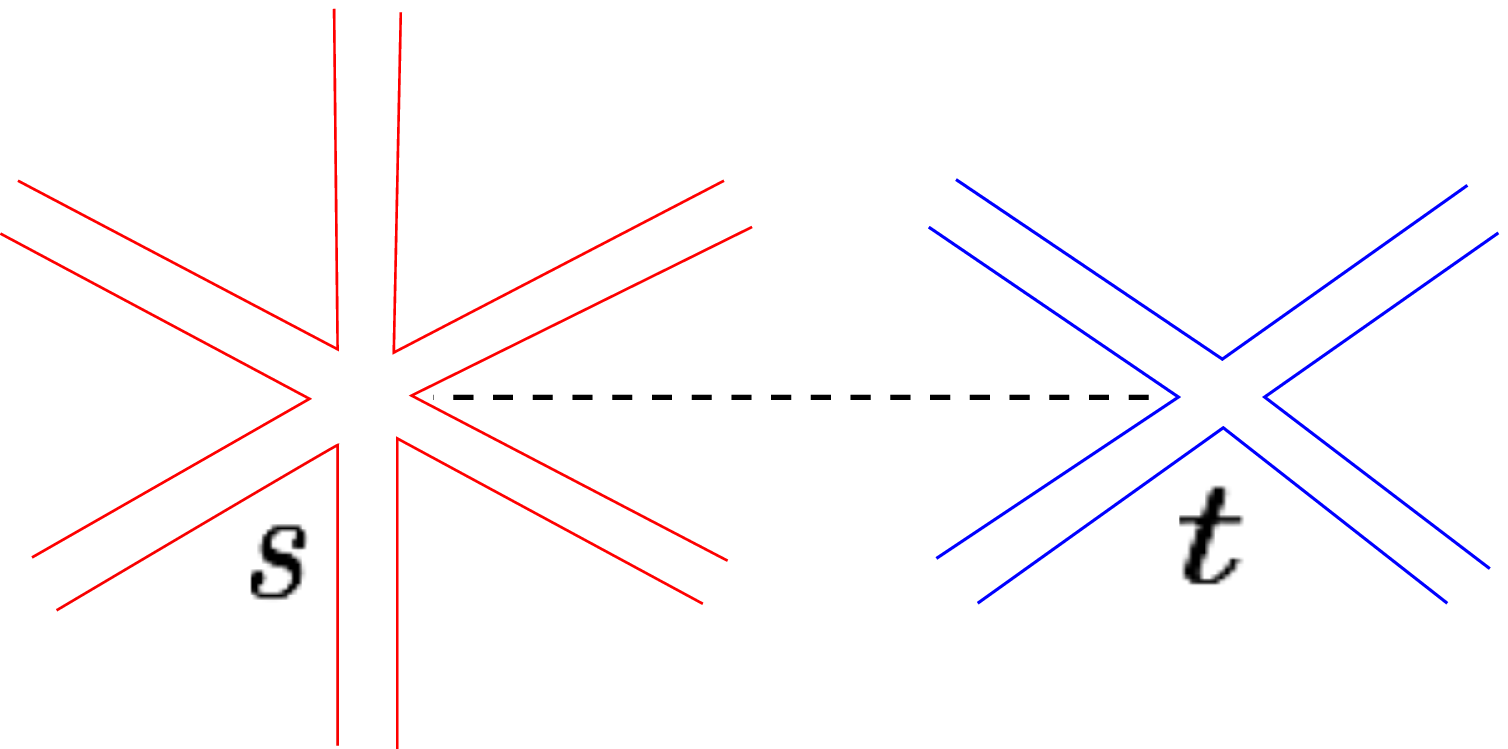}
\label{fig:s6r4}}\caption{ The red double line represents $\sigma$, 
and the blue double line represents $\rho$. 
The dashed line represents  a double trace interaction. 
A vertex in a single color, such as (c) or (d), represents an interaction in $U_{gauge}^r$,
 and thus $s\neq t$. 
A vertex in two colors, such as (e), represents an interaction in $U^r_{matter}$. }
\label{fig:Feynman rule for reduced model}
\end{center}
\end{figure}

From this action, we can read off the Feynman rule 
(see Figure \ref{fig:Feynman rule for reduced model}).
The propagators are given by
\bea
\langle \s_{sij} \s_{tkl}\rangle=\frac{t_1}{2N_1}\delta_{st}\delta_{il}\delta_{jk},\ \  \langle \rho_{s\alpha \beta} \rho_{t \gamma \delta }\rangle=-\frac{t_2}{2N_2}\delta_{st}\delta_{\alpha \delta}\delta_{\beta \gamma} .
\eea
The vertex of  $\tr  \s_s^a\ \tr \s_t^b$, or $\tr  \rho_s^a\ \tr \rho_t^b$, 
gives a factor
\bea
-\frac{2K_{ab}}{(s-t)^{a+b}}. \label{CS-type int}
\eea
The vertex of  $\tr  \s_s^a\ \tr \rho_t^b$ gives a factor
\bea
\frac{4K_{ab}}{2^{a+b}}
\biggl\{
\zeta\left(a+b, \frac{1}{4}+\frac{|s-t|}{2}\right)
-\zeta\left(a+b,\frac{3}{4}+\frac{|s-t|}{2}\right) \label{Matter-type int}
\biggr\}.
\eea
We write \eqref{CS-type int} and \eqref{Matter-type int} collectively as $V^{(a,b)}_{st}$;
\begin{equation}
V^{(a,b)}_{st}=
\begin{cases}
-2\frac{K_{ab}}{(s-t)^{a+b}} & \text{ for }V^{(a,b)}_{st} \in U_{gauge}^r.\\
\frac{4K_{ab}}{2^{a+b}}\biggl(\zeta\bigl(a+b, \frac{1}{4}+\frac{|s-t|}{2}\bigr)-\zeta\bigl(a+b,\frac{3}{4}+\frac{|s-t|}{2}\bigr)
\biggr) & \text{ for }V^{(a,b)}_{st} \in U_{matter}^r.
\end{cases}
\label{vertex in reduced model}
\end{equation}
Here, ``$V^{(a,b)}_{st} \in U_{gauge}^r$'' and ``$V^{(a,b)}_{st} \in U_{matter}^r$'' mean the vertices coming from the interactions $U_{gauge}^r$ 
and $U_{matter}^r$, respectively.

\begin{figure}[t]
\begin{center}
\subfigure[`tree' diagram ]{\includegraphics*[width=.30\linewidth]{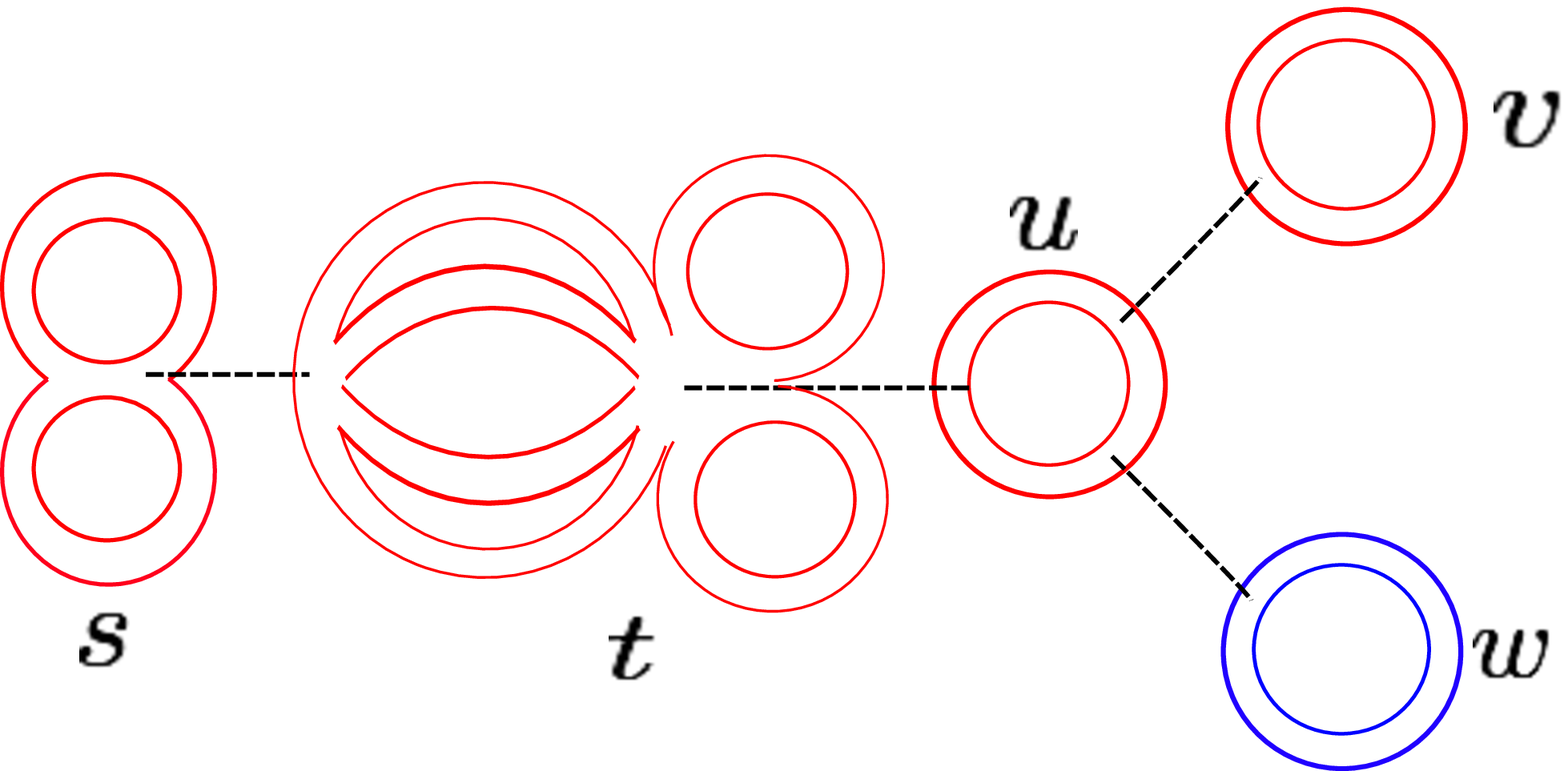}
\label{fig:tree}}
\subfigure[non-`tree' diagram]{\includegraphics*[width=.30\linewidth]{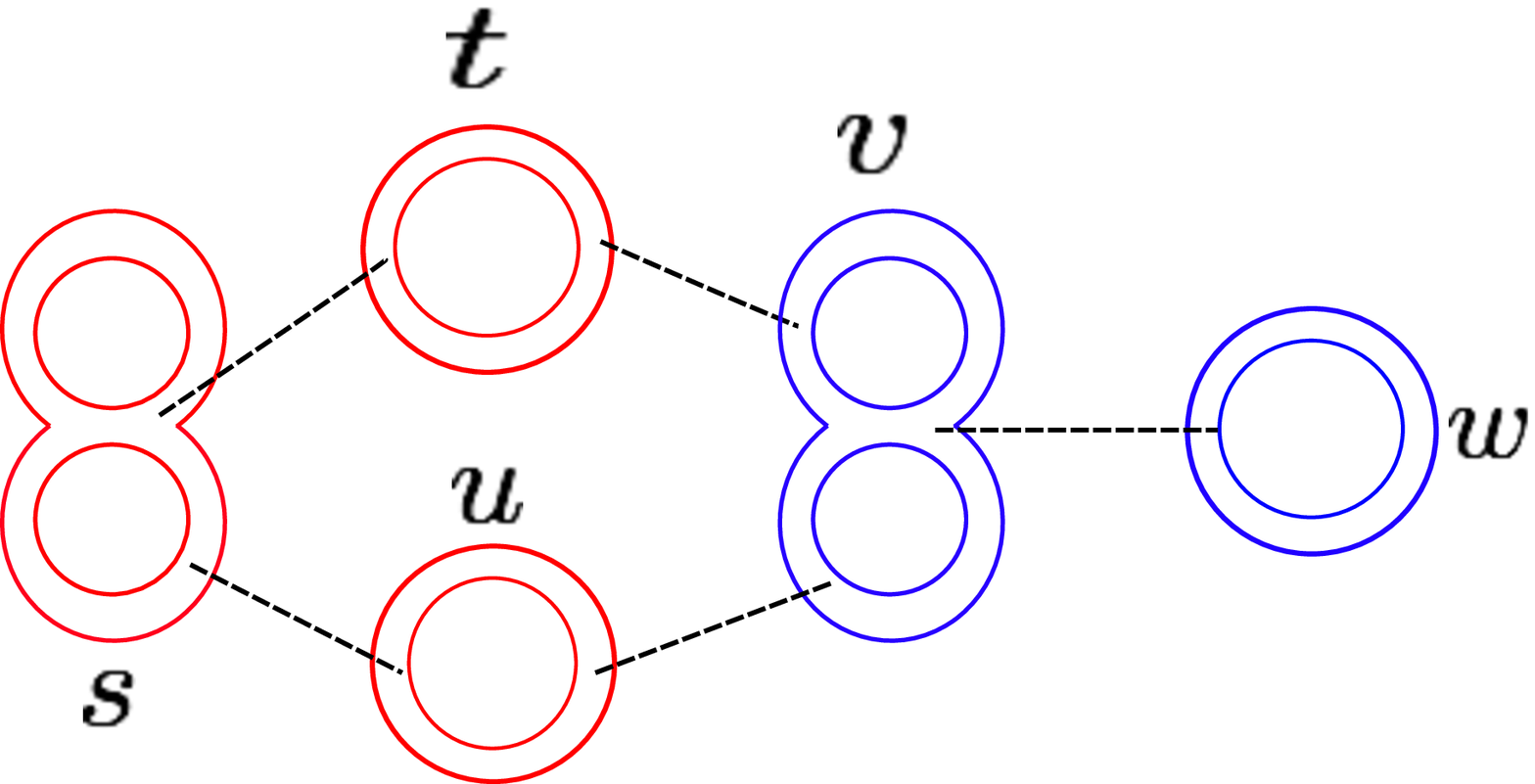}
\label{fig:loop}}
\caption{Examples of connected diagrams. While the dashed lines in (a) do not form a loop, the dashed lines in (b) do.  }
\label{fig:treevsloop}
\end{center}
\end{figure}

We discuss the calculation of the free energy based on the above Feynman rule. In this calculation, only connected diagrams are relevant as usual. Here, by ``connected'' we mean that any parts in a diagram are connected by dashed lines or by double lines. Figure  \ref{fig:treevsloop} shows examples of such diagrams. 
Since we are interested in the limit (\ref{planar limit}), 
let us consider what kind of diagrams contributes to the leading order 
of the $1/N_{1,2}$ expansion.  
It turns out that the leading contribution is given by the diagrams satisfying the following two conditions:

$Condition\ 1$. \; They are planar with respect to the double lines in the ordinary sense.

$Condition\ 2$. They can be separated into two parts by cutting any dashed lines.
We call a diagram satisfying the latter condition `tree' diagram since this condition is equivalent to that any dashed lines do not form a loop. We can check the latter condition explicitly for Figure \ref{fig:tree} as follows. Since $N_1$ and $N_2$ are in the same order in the limit (\ref{planar limit}), we denote the order of them collectively by $N$. Each propagator gives a factor of $N^{-1}$, each index loop gives $N$ and each vertex gives $N^{0}$. While (a) is proportional to $N^{-13}\times N^{15}=N^2$, (b) is proportional to $N^{-12}\times N^{12}=N^{0}$. Thus, Figure \ref{fig:loop} does not contribute in the limit (\ref{planar limit}).

\subsubsection{Feynman rule for  ABJM matrix model}
We next construct the Feynman rule for the ABJM matrix model on $S^3$, given by \eqref{2.5}.
This can be written as a manifestly $U(N_1) \times U(N_2)$ invariant form as follows,
\bea
\int d \s d \rho \exp\Biggl(-\frac{N_1}{\lambda_1}\tr \s^2 +\frac{N_2}{\lambda_2}\tr \rho^2 +{U}_{gauge} +{U}_{matter} \Biggr),\label{ ABJM matrix model action}
\eea
where we have defined the 't Hooft couplings $\lambda_i$ as $\frac{2\pi^2}{g_{s}} = \frac{N_i}{\lambda_i}$ ($i=1,2$), and $U_{gauge}$ and $U_{matter}$ are the double trace interactions, 
\bea
{U}_{gauge}&=& -\underset{a+b\in 2\IN}{\sum_{a,b}}{2 K_{ab}}\  \zeta(a+b) (\tr \s^a \ \tr \s^b + \tr \rho^a \ \tr \rho^b),\non\\
{U}_{matter}&=& \underset{a+b\in 2\IN}{\sum_{a,b}}{4K_{ab}\ \zeta\biggl(a+b, \frac{1}{2}\biggr)}\tr \s^a \tr \rho^b.\label{ ABJM int}
\eea

The vertices in the ABJM matrix model give the following factors,
\begin{equation}
{\mathcal V}^{(a,b)}\equiv
\begin{cases}
-4{K_{ab}}\zeta(a+b) &  \text{ for }\  {\mathcal V}^{(a,b)} \in  U_{gauge}\\
4{K_{ab}}\zeta(a+b,\frac{1}{2}) & \text{ for }\ {\mathcal V}^{(a,b)} \in U_{matter}.
\end{cases} \label{ ABJM vertex}
\end{equation}
The relevant diagrams in the limit (\ref{thooft lim in cs})
are planar and `tree' as in the case of the reduced model.

Compared with the Feynman rule in the reduced model, the ABJM matrix model does not have the indices $s, t, \cdots$ in the Feynman diagrams. 
We will show that after summing over these indices in the reduced model, each 
diagram in the reduced model reproduces the corresponding diagram in the  ABJM matrix model.

\subsubsection{Perturbative correspondence of free energy }
We compare the free energy of the reduced model with that of the ABJM theory. 
We will find that, in the limit (\ref{planar limit}), 
the free energy in the reduced model divided by $\Lambda+1$ coincides with 
that in the ABJM theory 
to all orders in the perturbation theory;
\bea
\frac{\mathcal{F}_{reduced}}{\Lambda+1}=\mathcal{F}_{ ABJM},
\label{equivalence of free energies}
\eea
 under the following identification of the coupling constants,
\bea
t_i = \lambda_i  \ ( i =1,2).
\label{coupling id}
\eea
We note that under \eqref{coupling id} the coefficients of the gaussian terms  of \eqref{reduced model action} and \eqref{ ABJM matrix model action} coincide and so  the factors coming from propagators in the reduced model and the ABJM theory  trivially agree for the same Feynman diagrams. Therefore,  in the following argument, we ignore the factors of  propagators and only take care of the factors coming from vertices in these  matrix models.

\begin{figure}[ t]
\begin{center}
\includegraphics[width=8cm]{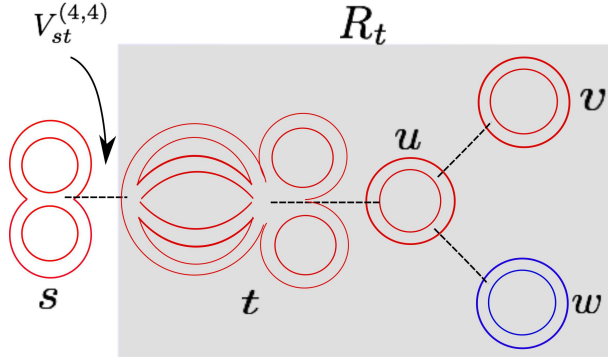}
\caption{This diagram has three outermost vertices, $V^{(4,4)}_{st}$, $V^{(2,2)}_{uv}$, $V^{(2,2)}_{uw}$.  Here, we have separated $V_{st}^{(4,4)}$ from the shaded part, $R_t$. }
\label{fig:Rt}
\end{center}
\end{figure}

In the reduced model, any `tree' diagram can be decomposed as
\bea
\sum_{s}\sum_t V^{(a,b)}_{st} R_t. \label{V and R}
\eea
Here, $V^{(a,b)}_{st}$ is an outermost vertex, that is, a vertex on the tip of a branch in the `tree' diagram, and $R_t$ represents the rest of the diagram. See Figure \ref{fig:Rt}. The explicit form of $R_t$ for the case shown in 
Figure \ref{fig:Rt} is given by \eqref{Rtexplicit}.  

\begin{figure}[ t]
\begin{center}
\includegraphics[width=5cm]{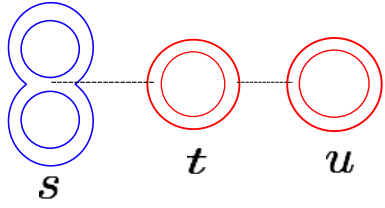}
\caption{The vertices in this diagram give a factor $\frac{1}{\Lambda+1}\sum_{s,t,u}V^{(2,4)}_{st}V^{(2,2)}_{tu}$, where $V^{(2,4)}_{st}\in U_{matter}^r$ and $V_{tu}^{(2,4)}\in U_{gauge}^r$. By \eqref{replacement}, in the $\Lambda\rightarrow \infty$ limit this is equal to ${\mathcal V}^{(2,4)}{\mathcal V}^{(2,2)}\frac{1}{\Lambda+1}\sum_u 1={\mathcal V}^{(2,4)}{\mathcal V}^{(2,2)}$, and thus we can recover the factor of the corresponding vertices in the ABJM theory.}
\label{fig:replacement}
\end{center}
\end{figure}

We will show that in the $\Lambda\rightarrow \infty$ limit, we can replace the sum $\sum_{s} V^{(a,b)}_{st}$ in \eqref{V and R} by the corresponding vertex in  the ABJM theory. That is,
\bea
\lim_{\Lambda\rightarrow \infty}\frac{1}{\Lambda+1} \sum_{t}\sum_s V_{st}^{(a,b)} R_t= \lim_{\Lambda\rightarrow \infty}\frac{{\mathcal V}^{(a,b)}}{\Lambda+1}\sum_t R_t, \label{replacement}
\eea
where $\mathcal{V}^{(a,b)}$ in the right-hand side is given by $\mathcal{V}^{(a,b)} \in U_{gauge}$ when ${V}_{st}^{(a,b)} \in U^r_{gauge}$ and  by $\mathcal{V}^{(a,b)} \in U_{matter}$ when ${V}_{st}^{(a,b)} \in U^r_{matter}$. 
If we establish \eqref{replacement}, by repeatedly replacing the vertices of the reduced model by those of the ABJM theory, the factor coming from  vertices in the reduced model agrees exactly with that in the Aharony-Bergman-Jafferis-Maldacena(ABJM)  matrix model. For an illustration of this procedure, see Figure \ref{fig:replacement}. Since 
this equivalence holds for all the `tree' diagrams,
we thus establish the perturbative equivalence of the free energy between the 
reduced model and the ABJM theory (\ref{equivalence of free energies}). 

We now prove \eqref{replacement}. For the vertex $V^{(a,b)}_{st}\in U_{matter}^r$,  the left-hand side in \eqref{replacement} is calculated as
\begin{align}
&\lim_{\Lambda\rightarrow \infty}\frac{1}{\Lambda+1} \sum_{t=-\halfL}^{\halfL} \sum_{\substack{s=-\halfL \\ (s\neq t)}}^{\halfL} \frac{-2K_{ab}}{(s-t)^{a+b}}R_t \non\\
&=\lim_{\Lambda\rightarrow \infty}\frac{-2}{\Lambda+1} \sum_{t=-\halfL}^{\halfL} K_{ab}R_t \Biggl(2\zeta(a+b)
-\zeta\biggl(a+b,\halfL+t+1\biggr)-\zeta\biggl(a+b,\halfL-t+1\biggr)\Biggr). \label{replacementproof}
\end{align}
The first term above agrees with the right-hand side in \eqref{replacement}. The remaining terms vanish in the $\Lambda\rightarrow \infty$ limit. To see this, we use the fact that $|R_t|$  has a $\Lambda$-independent upper bound denoted by $C$, which we prove in appendix \ref{sec:Rt}. 
By using this fact, the absolute value of the remaining terms is bounded from above by 
\bea
&&\lim_{\Lambda\rightarrow \infty}\frac{2|K_{ab}|C}{\Lambda+1}  \sum_{t=-\halfL}^{\halfL}\Biggl(\zeta\biggl(a+b,\halfL+t+1\biggr)+\zeta\biggl(a+b,\halfL-t+1\biggr)\Biggr)\non\\
&=&4 |K_{ab}|C \lim_{\Lambda\rightarrow \infty}\frac{1}{\Lambda+1}  \biggl( \sum_{n=0}^{\Lambda}\frac{1}{(n+1)^{a+b-1}}+\sum_{m=\Lambda+1}^{\infty}\frac{\Lambda+1}{(m+1)^{a+b}}\biggr)=0,\label{remaining part}
\eea
where we have used the definition of the generalized zeta function \eqref{zeta}. The terms in the second line are $\mathcal{O}(\log \Lambda/\Lambda)$ for $a+b=2$, and  $\mathcal{O}(1/\Lambda)$ for $a+b>2$. 

For the vertex $V^{(a,b)}_{st}\in U_{matter}^r$, the left-hand side in \eqref{replacement} is calculated as,
\begin{align}
&\frac{1}{\Lambda+1} \sum_{t=-\halfL}^{\halfL} \sum_{s=-\halfL}^{\halfL} \frac{4K_{ab}}{2^{a+b}}\Biggl(\zeta\biggl(a+b, \frac{1}{4}+\frac{|s-t|}{2}\biggr)-\zeta\biggl(a+b,\frac{3}{4}+\frac{|s-t|}{2}\biggr)
\Biggr)R_t \non\\
&=\frac{1}{\Lambda+1} \sum_{t=-\halfL}^{\halfL}  \frac{4K_{ab}}{2^{a+b}}
\Biggl(
2^{a+b}\zeta\biggl(a+b,\frac{1}{2}\biggr)
-\zeta\biggl(a+b,\frac{1}{4}+\frac{\halfL+t}{2}\biggr)-\zeta\biggl(a+b,\frac{3}{4}+\frac{\halfL-t}{2}\biggr)
\Biggr)R_t, \label{replacementproof2}
\end{align}
where we have used the equality, $\zeta(a+b,\frac{1}{4})+\zeta(a+b, \frac{3}{4})=2^{a+b}\zeta(a+b,\frac{1}{2})$.
The first term in \eqref{replacementproof2} agrees with the right-hand side in \eqref{replacement}. The remaining terms vanish in the $\Lambda\rightarrow \infty$ limit, which can be shown in the same way as \eqref{remaining part}.   Therefore, the equality \eqref{replacement} has been proved.

\begin{figure}[t]
\begin{center}
\subfigure[ ]{\includegraphics*[width=.30\linewidth]{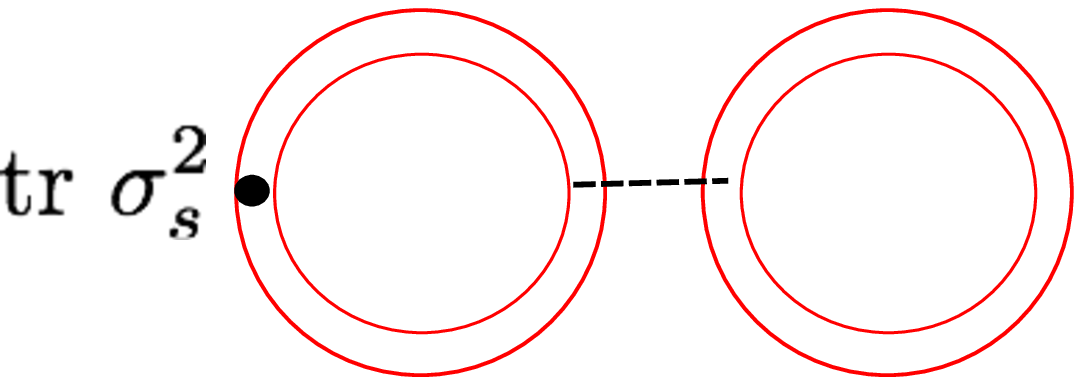}
\label{fig:boundary}}\ \  \ \ \ \ \ \ 
\subfigure[ ]{\includegraphics*[width=.350\linewidth]{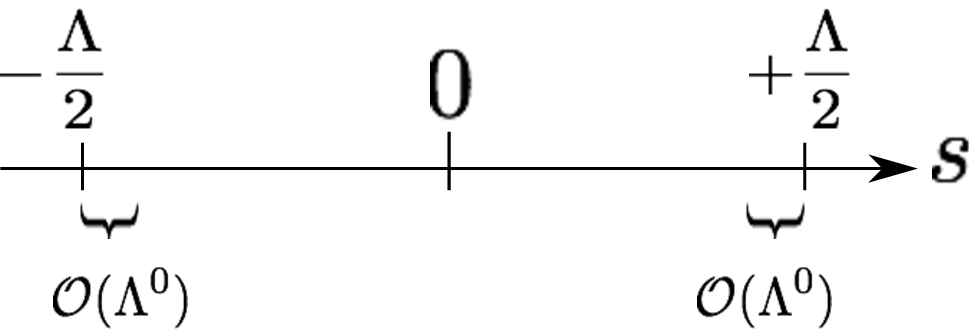}
\label{fig:boundary2}}
\caption{Figure (a): A diagram contributing to $\langle \tr \s_s^2 \rangle$. Figure (b) : Only $s$ near  $\pm \halfL$ feels the cutoff effect.}
\end{center}
\end{figure}

Here, we briefly comment on the cutoff effect of $\Lambda$. For this purpose, we consider the one-point function $\langle \tr \s_s^2 \rangle$  for a fixed $s \in \{-\halfL, \cdots, \halfL \}$.  Figure \ref{fig:boundary} is one of the diagrams which contributes to the one-point function, and leads to the following factor
\bea
\sum_{u (\neq s)} \frac{1}{(s-u)^4}
=\biggl( \sum_{n=1}^{\halfL-s}+\sum_{n=1}^{s+\halfL}\biggr)\frac{1}{n^4}. \label{bdry}
\eea
In the $\Lambda \rightarrow \infty$ limit, while \eqref{bdry} goes to $2\, \zeta(4)$
for all $s\in\{-\halfL+\mathcal{O}(\ln \Lambda), \cdots, \halfL-\mathcal{O}(\ln \Lambda)\}$,
it deviates from $2\, \zeta (4)$ for $s$ with $|s\pm \halfL| \sim \mathcal{O}(\Lambda^0)$.
Therefore, although \eqref{bdry} depends on the value of $s$, the dependence is negligible for almost all $s$ except near the cutoff  $\pm\frac{\Lambda}{2}$, and  only $s$ within $\mathcal{O}(\Lambda^0)$ from the boundaries feels the cutoff effect (See Figure \ref{fig:boundary2}). Note that this argument also holds for more general diagrams. Thus, the number of the modes affected by the cutoff is $\mathcal{O}(\Lambda^0)$, which is negligible compared to the total number, $\Lambda+1$. Since  the free energy and one-point functions of Wilson loops are written as an average value over various $s$,  the cutoff effect is negligible in the $\Lambda\rightarrow \infty$ limit.

\subsubsection{Perturbative correspondence of Wilson loop} 
We can easily prove the perturbative equivalence between the Wilson loop in the reduced model and that in the ABJM matrix model. While the Wilson loop in the ABJM theory is given by \eqref{2.8},   the corresponding object   in the reduced model is obtained by applying the representation \eqref{rep for sp} to 
(\ref{MMWilson}). Therefore, what we want to show is
\bea
\frac{1}{N_1}\langle \tr e^{2\pi \s}\rangle= \frac{1}{N_1 (\Lambda+1)}\sum_{s=-\halfL}^{\halfL} \langle \tr e^{2\pi \s_s}\rangle, \label{pert wilson}
\eea
where the $\langle \cdots \rangle$ means the average with respect to the eigenvalue integral of each theory.
In perturbation theory, both  of the left-hand side and the right-hand side above are calculated by expanding the exponentials. Thus, \eqref{pert wilson} holds if we have, for any positive integer $a$,
\bea
\langle \tr  \s^a \rangle= \frac{1}{\Lambda+1}\sum_{s=-\halfL}^{\halfL} \langle \tr \s^a_s\rangle.
\eea
We find that this is  true from \eqref{replacement}, and so is \eqref{pert wilson}.


\subsection{Large-$N$ equivalence of eigenvalue density}\label{equidensity}

In this section, we investigate the eigenvalue distributions of $\sigma_{s}$ and $\rho_s$ 
in the reduced model for the ABJM matrix model.
We show that if $s$ is sufficiently apart from the cutoff $\pm \Lambda/2$
the eigenvalue densities of $\sigma_s$ and $\rho_s$ coincide with those of $\sigma$ and $\rho$ in the ABJM matrix model, respectively,
and thereby prove the large-$N$ equivalence 
between the reduced model and the original ABJM theory.

We start with the ABJM matrix model (\ref{2.5}).
From (\ref{2.5}), the effective action for the eigenvalues are read off as
\begin{align}
S_{\mathrm{eff}}
&=\frac{N_1}{\lambda_1}\sum_i\sigma_i^2-\frac{N_2}{\lambda_2}\sum_\alpha \rho_\alpha^2
-\frac{1}{2}\sum_{i\neq j}\ln \sinh^2 \{\pi(\sigma_i-\sigma_j)\}
-\frac{1}{2}\sum_{\alpha\neq \beta}\ln \sinh^2 \{\pi(\rho_\alpha-\rho_\alpha)\} \n
&\quad
+2\sum_{i,\alpha}\ln\cosh\{\pi(\sigma_i-\rho_\alpha)\}.
\end{align}
Varying $S_{\mathrm{eff}}$ with respect to $\sigma_i$ and $\rho_\alpha$, we obtain 
 the saddle point equation
\begin{align}
0&=\frac{N_1}{\pi\lambda_1}\sigma_i
-\sum_{j(\neq i)}\coth\{\pi(\sigma_i-\sigma_j)\}
+\sum_{\alpha}\tanh\{\pi(\sigma_i-\rho_\alpha)\}, \n
0&=-\frac{N_2}{\pi\lambda_2}\rho_\alpha
-\sum_{\beta(\neq \alpha)}\coth\{\pi(\rho_\alpha-\rho_\beta)\}
-\sum_{i}\tanh\{\pi(\sigma_i-\rho_\alpha)\}.
\end{align}
Let us introduce the eigenvalue densities for $\sigma_i$ and $\rho_\alpha$ as
\begin{align}
\rho(x)=\frac{1}{N_1}\sum_{i=1}^{N_1}\delta(x-\sigma_{i}), \;\;\;
\tilde{\rho}(x)&=\frac{1}{N_2}\sum_{\alpha=1}^{N_2}\delta(x-\rho_{\alpha}).
\end{align}
Using these we can rewrite the saddle point equations as
\begin{align}
0=&\frac{1}{\pi\lambda_1} x 
-  \dashint dy \coth \{ \pi(x-y) \} \ \rho(y)
+ \frac{N_2}{N_1} \int dy \tanh \{ \pi(x-y) \} \ \tilde{\rho}(y), \n
0=&-\frac{1}{\pi\lambda_2} x 
-  \dashint dy \coth  \{\pi(x-y)\} \ \tilde{\rho}(y)
+ \frac{N_1}{N_2} \int dy \tanh \{ \pi(x-y)\} \ \rho(y),
\label{saddle point eq}
\end{align}
where $\dashint$ represents the Cauchy principal integral.
In the large-$N$ limit (\ref{thooft lim in cs}), 
the eigenvalues obeying \eqref{saddle point eq} form a continuous distribution
and so the eigenvalue densities become smooth functions.
The explicit solution of \eqref{saddle point eq} can be found in 
\cite{Drukker:2010nc,Marino:2011nm}.

Next let us consider our reduced model \eqref{sphere action} for the ABJM theory.
The effective action for the eigenvalues $\sigma_{si}$ and $\rho_{si}$ is given by
\begin{align}
S_{\mathrm{eff}}
&=\frac{N_1}{t_1}\sum_{si}\sigma_{si}^2-\frac{N_2}{t_2}\sum_{s\alpha}\rho_{s\alpha}^2 
-\frac{1}{2}\sum_{s}\sum_{i\neq j}\ln(\sigma_{si}-\sigma_{sj})^2
-\frac{1}{2}\sum_{s}\sum_{\alpha\neq \beta}\ln(\rho_{s\alpha}-\rho_{s\beta})^2
\n
&-\frac{1}{2}\sum_{s\neq t}\sum_{i,j} 
 \ln\left\{1+\frac{(\sigma_{si}-\sigma_{tj})^2}{(s-t)^2}\right\}
-\frac{1}{2}\sum_{s\neq t}\sum_{\alpha,\beta}
 \ln\left\{1+\frac{(\rho_{s\alpha}-\rho_{t\beta})^2}{(s-t)^2}\right\} \n
&-2\sum_{s,t}\sum_{i,\alpha}\sum_{J=|s-t|/2}^{\infty}
\left[\ln\left\{\left(2J+\frac{3}{2}\right)^2+(\sigma_{si}-\rho_{t\alpha})^2\right\}
-\ln\left\{\left(2J+\frac{1}{2}\right)^2+(\sigma_{si}-\rho_{t\alpha})^2\right\}
\right],
\end{align}
where the summation of $s,t$ is taken over $s,t=-\Lambda/2,\cdots, \Lambda/2$.
The saddle point equations are
\begin{align}
0&=\frac{N_1}{t_1}\sigma_{si}
-\sum_{j(\neq i)}\frac{1}{\sigma_{si}-\sigma_{sj}}
-\sum_{t(\neq s)}\sum_{j}\frac{\sigma_{si}-\sigma_{tj}}{(s-t)^2+(\sigma_{si}-\sigma_{tj})^2} \n
&-2\sum_{t,\alpha}\sum_{J=|s-t|/2}^{\infty}
\left[
\frac{\sigma_{si}-\rho_{t\alpha}}{\left(2J+\frac{3}{2}\right)^2+(\sigma_{si}-\rho_{t\alpha})^2}
-\frac{\sigma_{si}-\rho_{t\alpha}}{\left(2J+\frac{1}{2}\right)^2+(\sigma_{si}-\rho_{t\alpha})^2}
\right], \n
0&=-\frac{N_2}{t_2}\rho_{s\alpha}
-\sum_{\beta(\neq \alpha)}\frac{1}{\rho_{s\alpha}-\rho_{s\beta}}
-\sum_{t(\neq s)}\sum_{\beta}  
\frac{\rho_{s\alpha}-\rho_{s\beta}}{(s-t)^2+(\rho_{s\alpha}-\rho_{t\beta})^2} \n
&+2\sum_{t,i}\sum_{J=|s-t|/2}^{\infty}
\left[
\frac{\sigma_{ti}-\rho_{s\alpha}}{\left(2J+\frac{3}{2}\right)^2+(\sigma_{ti}-\rho_{s\alpha})^2}
-\frac{\sigma_{ti}-\rho_{s\alpha}}{\left(2J+\frac{1}{2}\right)^2+(\sigma_{ti}-\rho_{s\alpha})^2}
\right]. \label{saddle point eq reduced model 0}
\end{align}
We introduce the eigenvalue densities of $\sigma_s$ and $\rho_s$ 
in the reduced model as
\begin{align}
\rho^{[s]}(x)=\frac{1}{N_1}\sum_{i=1}^{N_1}\delta(x-\sigma_{si}), \;\;\;
\tilde{\rho}^{[s]}(x)&=\frac{1}{N_2}\sum_{\alpha=1}^{N_2}\delta(x-\rho_{s\alpha}),
\end{align}
and rewrite \eqref{saddle point eq reduced model 0} as
\begin{align}
0&=\frac{1}{t_1}x-\dashint dy \frac{1}{x-y}\rho^{[s]}(y)
-\sum_{t(\neq s)}\int dy
\frac{x-y}{(s-t)^2+(x-y)^2}\rho^{[t]}(y) \n
&-\frac{2N_2}{N_1}\sum_{t} \sum_{J=|s-t|/2}^{\infty} \int dy
\left[
\frac{x-y}{\left(2J+\frac{3}{2}\right)^2+(x-y)^2}
-\frac{x-y}{\left(2J+\frac{1}{2}\right)^2+(x-y)^2}
\right]\tilde{\rho}^{[t]}(y), \n
0&=-\frac{1}{t_2}x
-\dashint dy \frac{1}{x-y}\tilde{\rho}^{[s]}(y)
-\sum_{t(\neq s)}\int dy  
\frac{x-y}{(s-t)^2+(x-y)^2}\tilde{\rho}^{[t]}(y) \n
&-\frac{2N_1}{N_2}\sum_{t}\sum_{J=|s-t|/2}^{\infty}\int dy
\left[
\frac{x-y}{\left(2J+\frac{3}{2}\right)^2+(x-y)^2}
-\frac{x-y}{\left(2J+\frac{1}{2}\right)^2+(x-y)^2}
\right]\rho^{[t]}(y). \label{saddle point eq reduced model}
\end{align}
We can find a solution to these equations in the 
$\Lambda\rightarrow \infty$ limit as follows.
If one naively takes the $\Lambda\rightarrow \infty$ limit 
in \eqref{saddle point eq reduced model},
$(\rho^{[s]}(x),\tilde{\rho}^{[s]}(x)) = (\rho(x),\tilde{\rho}(x)) $
with $\lambda_1=t_1$ and $\lambda_2=t_2$
for arbitrary $s$ turns out to be a solution to \eqref{saddle point eq reduced model},
where $(\rho(x), \tilde{\rho}(x))$ is the solution to the saddle point 
equation 
\eqref{saddle point eq} of the ABJM matrix model.
This is because 
in this case \eqref{saddle point eq reduced model} reduces to \eqref{saddle point eq} 
(see appendix \ref{naive lambda limit}).
This solution represents infinitely many copies of that of the original ABJM matrix model.
Since the free energy and the Wilson loop in the reduced model are given 
by an average over all $s$'s as \eqref{equivalence of free energies} and \eqref{pert wilson},
they exactly coincide with those in the ABJM matrix model.

The densities $(\rho^{[s]}(x), \tilde{\rho}^{[s]}(x))$ with $s$ near the cutoff $\Lambda$
deviate from $(\rho(x), \tilde{\rho}(x))$.
This cutoff effect would spoil the above naive argument if the correlation range 
between $\rho^{[s]}$'s and $\tilde{\rho}^{[s]}$'s became larger as $\mathcal{O}(\Lambda)$.
In this case, the number of $(\rho^{[s]}(x), \tilde{\rho}^{[s]}(x))$ which deviates from 
$(\rho(x), \tilde{\rho}(x))$ would be $\mathcal{O}(\Lambda)$, 
namely, the number of the deviating densities and that of the densities coinciding 
with $(\rho(x), \tilde{\rho}(x))$ would become comparable.
Then the free energy and the Wilson loop in the reduced model 
given by the average over $s$'s would not coincide with those in the original model.

However, this is not the case of our reduced model.
It turns out that the correlation range is $\mathcal{O}(\Lambda^0)$,
and so the above naive argument is indeed valid.
In fact, in the saddle point equation for $\rho^{[s]}(x)$ (or $\tilde{\rho}^{[s]}(x)$), 
the coefficients of $\rho^{[t]}(x)$ and $\tilde{\rho}^{[t]}(x)$ are 
suppressed if $|t-s|$ is large enough.
As shown in appendix \ref{contributions from rhot}, 
the contributions from the terms with $|t-s|\geq \ln\Lambda$
 can be neglected in the $\Lambda\rightarrow \infty$ limit.
Namely, the profile of $\rho^{[s]}(x)$ (or $\tilde{\rho}^{[s]}(x)$) is determined 
only by $(\rho^{[t]}(x), \tilde{\rho}^{[t]}(x))$'s with $|t-s|\lesssim {\cal O}(\Lambda^0)$,
which means that the correlation range is 
sufficiently small compared to the system size $\Lambda$,
so that $(\rho^{[s]}(x), \tilde{\rho}^{[s]}(x))$ for $|s|\lesssim \Lambda/2-\ln \Lambda$ 
are not affected by the cutoff.
Therefore, in the $\Lambda\rightarrow \infty$ limit 
$(\rho^{[s]}(x), \tilde{\rho}^{[s]}(x))=(\rho(x), \tilde{\rho}(x))$ still holds 
except for very narrow region $\Lambda/2-\ln \Lambda\lesssim 
|s|\leq \Lambda/2$.
This is consistent with our observation in the perturbation theory 
mentioned in the last part of section 4.2.3.
Also this behavior of the densities is observed 
in the numerical simulation for the reduced model of the pure CS theory on $S^3$ 
\cite{Ishiki:2010pe}.
Although $(\rho^{[s]}(x), \tilde{\rho}^{[s]}(x))$ 
for $\Lambda/2-\ln \Lambda\lesssim |s|\leq \Lambda/2$ differs from 
$(\rho(x), \tilde{\rho}(x))$, their contributions to the physical quantities, 
such as the free energy and the BPS Wilson loops, are negligible
since the physical quantities are calculated as an average taken over 
all $s$'s.

Thus, for supersymmetric observables, 
the large-$N$ equivalence between the reduced model and the ABJM theory is also shown 
non-perturbatively through the saddle point method of the eigenvalue density.
One can also apply the saddle point analysis to 
general $\mathcal{N}=2$ non-chiral quiver CS theories and 
show the large-$N$ equivalence.

\section{Conclusion}
In this paper, we have studied the large-$N$ reduction for 
a general ${\cal N}=2$ non-chiral quiver CS theory on $S^3$.
We considered the reduced model of the
ABJM theory on $S^3$ as an illustration and
explained the calculation of the free energy and
the one-point function of the BPS Wilson loop operator
in the reduced model. 
We found that the localization technique reduces the calculation to eigenvalue integrals, 
as in the original theory on $S^3$.
To establish the large-$N$ equivalence, we first studied the integrals in the perturbation theory.
We constructed the Feynman rule from the eigenvalue integrals, 
and found that each Feynman diagram in the reduced model 
coincides with a corresponding diagram in the ABJM matrix model in
the continuum limit.
Hence, we conclude that these supersymmetric quantities are equivalent 
in two theories to all orders in the perturbative expansion.
Then we considered the saddle point configuration of the 
eigenvalues in the reduced model.
We found that in the continuum limit the cutoff effect is sufficiently small
and that the eigenvalue densities in the reduced model at the saddle point
consist of infinitely many copies of those in the original theory. 
This implies that 
the expectation values of supersymmetric observables in the reduced model,
which are written as the average over all the copies,
agree with those in the 
original theory in the continuum limit.
Thus the large-$N$ equivalence holds also non-perturbatively.
Our result gives a strong evidence that the non-perturbative formulation 
for supersymmetric theories based on the novel large-$N$ reduction 
works successfully.

\section*{Acknowledgements}
The work of G.I. and T.O. is supported by the Grant-in-Aid for the Global COE Program ``The Next Generation of Physics, Spun from Universality and Emergence'' from the Ministry of Education, Culture, Sports, Science and Technology (MEXT) of Japan. 
The work of S.S. is supported in part by the JSPS Research Fellowship for Young Scientists.

\appendix

\section{${\cal N}=2$ quiver CS theory on $S^3$}\label{ApC}
In this appendix, we summarize our convention for  
$\mathcal{N}=2$ quiver CS theory on $S^3$ \cite{Hama:2010av}.
We consider the gauge vector multiplet and the  
matter chiral multiplets in the adjoint and in the bifundamental 
representation.

A gauge multiplet contains fermionic (Grassmaniann) fields 
$\{\lambda, \bar{\lambda} \}$ 
as well as bosonic fields $\{A_\mu,\sigma ,D\}$.
There are two kinds of supersymmetric action for this 
multiplet: the CS action and the YM action, which are defined by 
\begin{align}
&S_{CS}=-\int \dif{^3x}\; \tr \!\! \left[ \varepsilon ^{\mu \nu \lambda}(A_\mu \partial _\nu A_\lambda -\frac{2i}{3}A_\mu A_\nu A_\lambda )+\sqrt{g}( -\bar \lambda \lambda +2D\sigma ) \right], \label{scs} \\
&S_{YM}=\int d^3x \sqrt{g}\; \tr \Big[ \frac{1}{4}F^{\mu \nu}F_{\mu \nu}+\frac{1}{2}(\sigma +D)^2+\frac{1}{2}(D_\mu \sigma )^2+\frac{i}{2}\bar \lambda \gamma ^\mu D_\mu \lambda -\frac{1}{4}\bar \lambda \lambda +\frac{i}{2}\bar \lambda [\sigma ,\lambda ]\Big].
\label{sym}
\end{align}
The field strength is defined as usual by
$F_{\mu \nu}=\partial _\mu A_\nu-\partial _\nu A_\mu-i[A_\mu ,A_\nu ]$, and
the covariant derivative contains the gauge and the spin connections,
\begin{align}
D_\mu \lambda =\partial _\mu \lambda 
+\frac{1}{4}\omega _{\mu}^{\;\; bc}\gamma _{bc}\lambda 
-i[ A_\mu , \lambda ].
\end{align}

A matter multiplet in the bifundamental representation 
consists of bosonic fields $\{\phi, \bar\phi , F, \bar F \}$  
and fermionic fields $\{\psi, \bar\psi \}$.
We assume that they couple to a gauge multiplet 
$\{A_\mu ,\lambda ,\sigma ,D\}$ as the fundamental representation and 
to another gauge multiplet $\{ B_\mu ,\eta ,\rho ,\tilde D\}$
as the anti-fundamental.
The supersymmetric action for this multiplet is given 
by\footnote{In general, the theory may have a superpotential.
We ignore it in this paper 
since it is irrelevant for the localization calculation.} 
\begin{align}
&S_{matter}=\int d^3x \sqrt{g}\; \tr [D^\mu \bar \phi D_\mu \phi -i\bar \psi \gamma ^\mu D_\mu \psi +q(2-q)\bar \phi \phi -\frac{2q-1}{2}\bar \psi \psi +i(2q-1)\bar \phi \nabla (\sigma ,\rho )\phi \nonumber \\
&\; +i\bar \psi \nabla (\sigma ,\rho ) \psi +i\bar \psi \nabla (\lambda ,\eta  )\phi -i\bar \phi \nabla (\bar \lambda ,\bar{\eta })\psi +i\bar \phi \nabla (D,\tilde D)\phi +\bar \phi \nabla (\sigma ,\rho )^2\phi +\bar FF].
\label{smatt}
\end{align}
Here, $q$ is the anomalous dimension and
$\nabla (A,B)$ is defined as the operator which acts as
\begin{equation}
\nabla (A,B)\phi :=A\phi -\phi B, \;\; 
\nabla (A,B)\bar \phi :=B\bar \phi -\bar \phi A
\label{nabla}
\end{equation}
on the bifundamental and on the anti-bifundamental field, respectively.
The covariant derivative acts on the spinors as
\begin{align}
D_\mu \psi =\partial _\mu \psi +\frac{1}{4}\omega _{\mu}^{\;\; bc}\gamma _{bc}\psi -i\nabla (A_\mu ,B_\mu )\psi.
\end{align}

The actions (\ref{scs}),(\ref{sym}) and (\ref{smatt}) are
invariant under the supersymmetry transformations,
\begin{align}
\delta A_\mu =&\frac{i}{2}(\bar \lambda \gamma _\mu \epsilon -\bar \epsilon \gamma _\mu \lambda ),\nonumber\\
\delta \sigma =&-\frac{1}{2}(\bar \lambda \epsilon -\bar \epsilon \lambda ),\nonumber\\
\delta \lambda =&\frac{1}{2}\gamma ^{\mu \nu}\epsilon F_{\mu \nu}-D\epsilon +i\gamma ^\mu \epsilon D_\mu \sigma +\frac{2i}{3}\sigma \gamma ^\mu D_\mu \epsilon ,\nonumber\\
\delta \bar \lambda =&\frac{1}{2}\gamma ^{\mu \nu}\bar \epsilon F_{\mu \nu}+D\bar \epsilon -i\gamma ^\mu \bar \epsilon D_\mu \sigma -\frac{2i}{3}\sigma \gamma ^\mu D_\mu \bar \epsilon ,\nonumber\\
\delta D=&-\frac{i}{2}D_\mu \bar \lambda \gamma ^\mu  \epsilon -\frac{i}{2}\bar \epsilon \gamma ^\mu  D_\mu \lambda +\frac{i}{2}[\bar \lambda \epsilon ,\sigma ]+\frac{i}{2}[\bar \epsilon \lambda ,\sigma ]-\frac{i}{6}(\bar \lambda \gamma ^\mu D_\mu \epsilon +D_\mu \bar \epsilon \gamma ^\mu \lambda ),
\label{susy for gauge on s3}
\end{align}
for the gauge multiplet and
\begin{align}
\delta \phi =&\bar \epsilon \psi ,\nonumber\\
\delta \bar \phi =&\epsilon \bar \psi ,\nonumber\\
\delta \psi =&i\gamma ^\mu  \epsilon D_\mu \phi +i\epsilon \nabla (\sigma ,\rho )\phi +\frac{2iq}{3}\gamma ^\mu D_\mu \epsilon \phi +\bar \epsilon F ,\nonumber\\
\delta \bar \psi =&i\gamma ^\mu \bar \epsilon D_\mu \bar \phi -i\nabla (\sigma ,\rho )\bar \phi \bar \epsilon +\frac{2iq}{3}\bar \phi \gamma ^\mu D_\mu \bar \epsilon +\bar F\epsilon ,\nonumber\\
\delta F=&\epsilon (i\gamma ^\mu D_\mu \psi -i\nabla (\sigma ,\rho )\psi -i\nabla (\lambda ,\eta )\phi )+\frac{i}{3}(2q-1)D_\mu \epsilon \gamma ^\mu \psi ,\nonumber\\
\delta \bar F=&\bar \epsilon (i\gamma ^\mu D_\mu \bar \psi +i\nabla (\sigma ,\rho )\bar \psi -i\nabla (\bar \lambda ,\bar \eta )\bar \phi )+\frac{i}{3}(2q-1)D_\mu \bar \epsilon \gamma ^\mu \bar \psi ,
\label{susy for matter on s3}
\end{align}
for the bifundamental matter multiplet.
The Grassmaniann parameters $\epsilon$ and $\bar \epsilon$ satisfy 
the Killing spinor equation,
\begin{equation}
D_\mu \epsilon=\pm \frac{i}{2}\gamma_{\mu} \epsilon.
\label{KS eq}
\end{equation}
In the right invariant frame defined in appendix \ref{s3},
solutions to the Killing spinor equation are given by
\begin{align}
&\epsilon =\epsilon _0 \;\; {\rm and} \;\; \epsilon= g\epsilon _0
\label{solution to KSE}
\end{align}
for the upper and the lower sign in (\ref{KS eq}), respectively, 
where $\epsilon_0$ is a constant spinor on $S^3$ and $g$ is 
a group element of $SU(2)$ defined in (\ref{Euler angles}).

The supersymmetry transformation $\delta$ can be divided into two parts 
generated by $\epsilon$ and $\bar \epsilon$
as $\delta =\delta _\epsilon +\delta _{\bar \epsilon}$.
While two unbarred or two barred supersymmetries commute,
the commutator $[\delta _\epsilon ,\delta _{\bar \epsilon}]$ is 
given by a sum of translation, gauge transformation, 
Lorentz rotation, dilatation, and R-rotation.

One can also obtain the action and the supersymmetry transformation 
for an adjoint matter multiplet by
identifying one gauge multiplet with the other in 
(\ref{smatt}) and (\ref{susy for matter on s3}).

\section{Our convention for $S^3$}\label{s3}
In this appendix, we summarize our convention for $S^3$ with a unit 
radius (see also \cite{Ishii:2008tm,Ishii:2008ib}).
$S^3$ is viewed as the $SU(2)$ group manifold. We parametrize an element
of $SU(2)$ in terms of the Euler angles as
\begin{equation}
g=e^{-i\varphi \gamma_3/2}e^{-i\theta \gamma_2/2}e^{-i\psi \gamma_3/2},
\label{Euler angles}
\end{equation}
where $0\leq \theta\leq \pi$, $0\leq \varphi < 2\pi$, $0\leq \psi < 4\pi$ 
and $\gamma_a$ are the Pauli matrices.
The periodicity for these angle variables is given by
\begin{align}
(\theta,\varphi,\psi)\sim (\theta,\varphi+2\pi,\psi+2\pi)\sim (\theta,\varphi,\psi+4\pi).
\label{periodicity on S^3}
\end{align}

The isometry of $S^3$ corresponds to the left and the right multiplications
of $SU(2)$ elements on $g$. 
We construct the right-invariant 1-forms under the multiplications,
\begin{equation}
dgg^{-1}=-i e^a \gamma_a.
\label{ri 1-form}
\end{equation}
The explicit form of $e^a$ is given by
\begin{eqnarray}
&&e^1=\frac{1}{2}(-\sin \varphi d\theta + \sin\theta\cos\varphi d\psi),\nonumber\\
&&e^2=\frac{1}{2}(\cos \varphi d\theta + \sin\theta\sin\varphi
 d\psi),\nonumber\\
&&e^3=\frac{1}{2}(d\varphi + \cos\theta d\psi).
\label{ri explicit}
\end{eqnarray} 
It is easy to see that $e_a$ satisfy the Maurer-Cartan equation,
\begin{equation}
de^a-\varepsilon_{abc}e^b\wedge e^c=0.
\label{Maurer-Cartan}
\end{equation}
We take $e^a$ as the vielbein in this paper. In this frame, 
the spin connection is simply given by $\omega^{ab}=\varepsilon^{abc}e^c$.
The metric is given by
\begin{equation}
ds^2=e^ae^a=\frac{1}{4}\left(
d\theta^2+\sin^2\theta d\varphi^2 +(d\psi+\cos\theta d\varphi)^2\right).
\label{metric of S^3}
\end{equation}


\section{Commutator between $\delta _{\epsilon}$ and $\delta _{\bar \epsilon}$ in reduced model}\label{A.comm}
The actions of $[\delta _\epsilon ,\delta _{\bar \epsilon}]$ on 
all the matrices are shown below:
\begin{align}
[\delta _\epsilon ,\delta _{\bar \epsilon}]A_a=&\Theta _a^{\;\; b}A_b+i[\chi ,A_a],\nonumber \\
[\delta _\epsilon ,\delta _{\bar \epsilon}]\sigma =&i[\chi ,\sigma ],\nonumber \\
[\delta _\epsilon ,\delta _{\bar \epsilon}]\lambda =&\frac{1}{4}\Theta _{ab}\gamma ^{ab}\lambda +i[\chi ,\lambda ]+\alpha \lambda ,\nonumber \\
[\delta _\epsilon ,\delta _{\bar \epsilon}]\bar \lambda =&\frac{1}{4}\Theta _{ab}\gamma ^{ab}\bar \lambda +i[\chi ,\bar \lambda ]-\alpha \bar \lambda ,\nonumber \\
[\delta _\epsilon ,\delta _{\bar \epsilon}]D=&i[\chi ,D],
\end{align}
and
\begin{align}
[\delta _\epsilon ,\delta _{\bar \epsilon}]\phi =&i\nabla (\chi ,\tilde \chi )\phi -q\alpha \phi ,\nonumber \\
[\delta _\epsilon ,\delta _{\bar \epsilon}]\phi =&i\nabla (\chi ,\tilde \chi )\bar \phi +q\alpha \bar \phi ,\nonumber \\
[\delta _\epsilon ,\delta _{\bar \epsilon}]\psi =&\frac{1}{4}\Theta _{ab}\gamma ^{ab}\psi +i\nabla (\chi ,\tilde \chi )\psi +(1-q)\alpha \psi ,\nonumber \\
[\delta _\epsilon ,\delta _{\bar \epsilon}]\bar \psi =&\frac{1}{4}\Theta _{ab}\gamma ^{ab}\bar \psi +i\nabla (\chi ,\tilde \chi )\bar \psi -(1-q)\alpha \bar \psi ,\nonumber \\
[\delta _\epsilon ,\delta _{\bar \epsilon}]F=&i\nabla (\chi ,\tilde \chi )F+(2-q)\alpha F,\nonumber \\
[\delta _\epsilon ,\delta _{\bar \epsilon}]\bar F=&i\nabla (\chi ,\tilde \chi )\bar F-(2-q)\alpha \bar F,
\end{align}
where
\begin{align}
\Theta ^{ab}:=&2i\varepsilon ^{abc}\bar \epsilon \gamma _c\epsilon ,\nonumber \\
\chi :=&-iA_a\bar \epsilon \gamma ^a\epsilon +\sigma \bar \epsilon \epsilon \nonumber \\
\tilde \chi :=&-iB_a\bar \epsilon \gamma ^a\epsilon +\rho \bar \epsilon \epsilon \nonumber \\
\alpha :=&\bar \epsilon \epsilon .
\end{align}
The action of $[\delta _\epsilon ,\delta _{\bar \epsilon}]$ on the gauge multiplet $\{ B_a,\rho ,\eta ,\tilde D\}$ takes the same form as
$\{ A_a,\sigma ,\lambda ,D\}$.
We can read off from the above equations that $\Theta ^{ab}$ are parameters of $SU(2)$ rotation, $\chi$ and $\tilde \chi$ are gauge transformations for $A_a$ and $B_a$, respectively, and $\alpha$ is R-rotation.

\section{Fuzzy spherical harmonics}\label{A.harmonics}
In this appendix, we review the fuzzy spherical harmonics which
form a basis of rectangular matrices \cite{Ishii:2008ib,Ishiki:2006yr}.

Let us consider a $(2j_s+1)\times (2j_t+1)$ rectangular complex matrix.
Such a matrix $M^{(s,t)}$ can be generally expanded as
\begin{equation}
M^{(s,t)}=\sum_{m_s,m_t}M_{m_sm_t}|j_sm_s\rangle \langle j_tm_t |,
\end{equation}
by using a basis $\{ |jm \rangle \;|m=-j,-j+1,\cdots,j \}$ 
of the spin $j$ representation space 
of $SU(2)$ algebra. 
We define an operation which multiplies the representation 
matrices of the $SU(2)$ generators from left and right:
\begin{align}
L_a\circ M^{(s,t)}&=\sum_{m_s,m_t}M_{m_sm_t}(L_a^{[j_s]}|j_sm_s\rangle \langle j_tm_t |-|j_sm_s\rangle \langle j_tm_t |L_a^{[j_t]}),
\label{LcM}
\end{align}
where $L_a^{[j]}$ stands for the spin $j$ representation matrix of the 
generator.

We can construct another basis of the rectangular matrices 
denoted by $\{ \hat Y_{Jm(j_sj_t)} \}$ such that they satisfy
\begin{align}
(L_a\circ )^2\hat Y_{Jm(j_sj_t)}=&J(J+1)\hat Y_{Jm(j_s,j_t)},\nonumber \\
L_\pm \circ \hat Y_{Jm(j_sj_t)}=&\sqrt{(J\mp m)(J\pm m+1)}\hat Y_{Jm\pm 1(j_s,j_t)},\nonumber \\
L_3\circ \hat Y_{Jm(j_sj_t)}=&m\hat Y_{Jm(j_s,j_t)}.
\end{align}
$ \hat Y_{Jm(j_sj_t)}$ are
called scalar fuzzy spherical harmonics and defined by
\begin{equation}
\hat Y_{Jm(j_sj_t)}=
\sum_{m_s,m_t}(-)^{-j_s+m_t}C^{Jm}_{j_sm_sj_tm_t}|j_sm_s\rangle 
\langle j_tm_t|,
\end{equation}
where $C^{Jm}_{j_sm_sj_tm_t}$ are the 
Clebsch-Gordan coefficients.
Their hermitian conjugates are given by
\begin{equation}
(\hat Y_{Jm(j_sj_t)})^\dagger =(-)^{m-(j_s-j_t)}\hat Y_{J-m(j_tj_s)},
\end{equation}
and they satisfy the orthogonality relation
\begin{equation}
\tr \left\{ (\hat Y_{Jm(j_sj_t)})^\dagger 
\hat Y_{J^\prime m^\prime (j_s^\prime j_t^\prime )}\right\} =\delta _{J,J^\prime}\delta _{m,m^\prime}.
\end{equation}

Then we define the vector fuzzy spherical harmonics 
$\hat Y_{Jm(j_sj_t)a}^{\rho}$ and the 
spinor fuzzy spherical harmonics $\hat Y_{Jm(j_s,j_t)\alpha}^{\kappa}$,
where $\rho =-1,0,1$, $\kappa =-1,1$. The indices 
$a=1,2,3$ and $\alpha=1,2$ are those for vectors and spinors, 
respectively\footnote{Here, we mean just a set of three or two matrices
by `vector' or `spinor'. This terminology makes sense only when 
we regard them as the regularized version of the vector and the spinor
spherical harmonics on $S^2$ in the presence of a monopole.
See \cite{Ishiki:2006yr} and references therein.}.
They are written in terms of the scalar fuzzy spherical harmonics,
\begin{align}
\hat Y_{Jm(j_sj_t)a}^\rho =&i^\rho \sum_{n,p}V_{an}C^{Qm}_{\tilde Qp1n}\hat Y_{\tilde Qp(j_sj_t)},
\label{vector harmonics}
\\
\hat Y_{Jm(j_sj_t)\alpha}^\kappa =&\sum_{p}C^{Um}_{\tilde Up\frac{1}{2}\alpha}\hat Y_{\tilde Up(j_sj_t)},
\label{spinor harmonics}
\end{align}
where $Q=J+\delta _{\rho ,1}, \tilde Q=J+\delta _{\rho ,-1}, U=J+\frac{1}{2}\delta _{\kappa ,1}$ and $\tilde U=J+\frac{1}{2}\delta _{\kappa ,-1}$. 
$V$ is an unitary matrix defined by
\begin{align}
V=\frac{1}{\sqrt{2}}
\begin{pmatrix}
-1&0&1\\
-i&0&-i\\
0&\sqrt{2}&0
\end{pmatrix}.
\end{align}
The vector and the spinor harmonics satisfy the following formulae,
\begin{align}
&L_a\circ \hat Y_{Jm(j_sj_t)a}^\rho =\sqrt{J(J+1)}\delta _{\rho ,0}\hat Y_{Jm(j_sj_t)},
\label{LYvec} \\
&i\varepsilon _{abc}L_b\circ \hat Y_{Jm(j_sj_t)c}^\rho 
+\hat Y_{Jm(j_s,j_t)a}^\rho =\rho (J+1)\hat Y_{Jm(j_sj_t)a}^\rho
\label{LY+Yvec}
\\
&(\gamma ^a)_\alpha ^{\;\; \beta}L_a\circ 
\hat Y^\kappa _{Jm(j_sj_t)\beta}+\frac{3}{4}
\hat Y^\kappa _{Jm(j_sj_t)\alpha}=\kappa ( J+\frac{3}{4}) 
\hat Y^\kappa _{Jm(j_sj_t) \alpha}.
\label{LY+Yspn}
\end{align}
Their hermitian conjugates are given by
\begin{align}
(\hat Y_{Jm(j_sj_t)a}^\rho )^\dagger =&(-)^{m-(j_s-j_t)+1}\hat Y_{J-m(j_tj_s)a}^\rho ,
\label{YvecHC}\\
(\hat Y_{Jm(j_sj_t)\alpha}^\kappa )^\dagger =&(-)^{m-(j_s-j_t)+\kappa \alpha +1}\hat Y_{J-m(j_tj_s)-\alpha}^\kappa .
\label{YspnHC}
\end{align}
They also satisfy the orthogonality relations,
\begin{align}
\tr \left\{ (\hat Y_{Jm(j_sj_t)a}^\rho )^\dagger 
\hat Y^{\rho ^\prime}_{J^\prime m^\prime (j_s^\prime j_t^\prime )a}\right\} =\delta _{J,J^\prime}\delta _{m,m^\prime}\delta _{\rho ,\rho ^\prime},
\label{trYvec}\\
\tr \left\{ (\hat Y_{Jm(j_sj_t)\alpha}^\kappa )^\dagger 
\hat Y^{\kappa ^\prime}_{J^\prime m^\prime (j_s^\prime j_t^\prime )\alpha }\right\} =\delta _{J,J^\prime}\delta _{m,m^\prime}\delta _{\kappa ,\kappa ^\prime}.
\label{trYspn}
\end{align}

\section{Finiteness of $R_t$ in the limit \eqref{V and R}} \label{sec:Rt}
We prove the finiteness of $R_t$  in the limit of $\Lambda\rightarrow \infty$.
We first give the proof for  a simple case shown in Figure \ref{fig:Rt}.   

 In this case, $R_t$ is given by
\bea
R_t= \sum_{\substack{u=-\halfL \\ (u\neq t)}}^{\halfL} \sum_{\substack{v=-\halfL \\ (v\neq u)}}^{\halfL} \sum_{w=-\halfL}^{\halfL}V_{tu}^{(8,2)}V_{uv}^{(2,2)}V_{uw}^{(2,2)},\label{Rtexplicit}
\eea
and using the explicit form \eqref{vertex in reduced model}, this becomes
\bea
\sum_{\substack{u=-\halfL \\ (u\neq t)}}^{\halfL} \sum_{\substack{v=-\halfL \\ (v\neq u)}}^{\halfL} \sum_{w=-\halfL}^{\halfL}\frac{1}{(t-u)^{10}}\frac{1}{(u-v)^4} \frac{1}{2^4}\Biggl(\zeta\biggl(4, \frac{1}{4}+\frac{|u-w|}{2}\biggr)-\zeta\biggl(4,\frac{3}{4}+\frac{|u-w|}{2}\biggr)\Biggr), \label{Rt example}
\eea
where we have omitted the $\Lambda$-independent factor, $K_{8\, 2} K_{2\, 2} K_{2\, 2}$, defined in \eqref{Kab}.
To evaluate this in the limit $\Lambda\rightarrow \infty$, we make use of the following inequalities,
\begin{align}
&\sum_{\substack{s=-\halfL \\ (s\neq t)}}^{\halfL}\frac{1}{(s-t)^{2l}}= 2\zeta(z)-  \zeta\biggl(z,\halfL+s+1\biggr)-\zeta\biggl(z, \halfL-s+1\biggr)<2\zeta(z),\non\\
&\sum_{s=-\halfL}^{\halfL}\frac{1}{2^z}\Biggl(\zeta\biggl(z, \frac{1}{4}+\frac{|s-t|}{2}\biggr)-\zeta\biggl(z,\frac{3}{4}+\frac{|s-t|}{2}\biggr)\Biggr) \non\\&\ \ \ \ \ \ \ \ \ =2^z \zeta\biggl(z,\frac{1}{2}\biggr) -\zeta\biggl(z,\frac{3}{4}+\frac{t+\halfL}{2}\biggr)-\zeta\biggl(z,\frac{3}{4}+\frac{\halfL-t}{2}\biggr)<2^z \zeta\biggl(z,\frac{1}{2}\biggr).
\end{align}
By utilizing these, \eqref{Rt example} is bounded from above by 
\bea
2\zeta(10) \times 2\zeta(4) \times \zeta\left(4,\frac{1}{2}\right),
\eea
and thus $R_t$ is finite even for  $\Lambda\rightarrow \infty$.

As is obvious from this proof,  a general $R_t$   can also be bounded from above by  a  product of $\zeta$ functions, and thus we complete the proof of  the finiteness of $R_t$.

\section{Saddle point equation in reduced model}

\subsection{Naive $\Lambda\rightarrow \infty$ limit \label{naive lambda limit}}

In this appendix, we show that, 
if one naively takes the $\Lambda\rightarrow \infty$ limit 
in \eqref{saddle point eq reduced model},
 $\rho^{[s]}=\rho$ and $\tilde{\rho}^{[s]}=\tilde{\rho}$ for all $s$ is a solution of 
 \eqref{saddle point eq reduced model}.
To see this, we take the $\Lambda\rightarrow \infty$ limit 
and substitute the ansatz $\rho^{[s]}=\hat{\rho}$ and $\tilde{\rho}^{[s]}=\hat{\tilde{\rho}}$ 
for all $s$ into \eqref{saddle point eq reduced model} 
(Here we show only the first equation of \eqref{saddle point eq reduced model} for simplicity.
We can show the second one completely in the same manner.), 
then we obtain
\begin{align}
0=&\frac{1}{t_1}x
-\sum_{t=-\infty}^{\infty}\dashint dy \frac{x-y}{(s-t)^2+(x-y)^2}\hat{\rho}(y)
 \n
&-\frac{2N_2}{N_1}\sum_{t=-\infty}^{\infty}\sum_{J=\frac{|s-t|}{2}}^{\infty}
\int dy \left\{
\frac{x-y}{(2J+\frac{3}{2})^2+(x-y)^2}-\frac{x-y}{(2J+\frac{1}{2})^2+(x-y)^2} \right\}
\hat{\tilde{\rho}}(y). 
\label{saddle point eq reduced model 2}
\end{align}
By using the following formulae
\begin{align}
\coth x&=\frac{1}{x}+\sum_{n=1}^{\infty}\frac{2x}{\pi^2n^2+x^2}
        =\sum_{n=-\infty}^{\infty}\frac{x}{\pi^2n^2+x^2}, \n
\tanh x&=\sum_{n=1}^{\infty}\frac{2x}{\pi^2(n-\frac{1}{2})^2+x^2}
        =\sum_{n=-\infty}^{\infty}\frac{x}{\pi^2(n-\frac{1}{2})^2+x^2},
\end{align}
it is easily seen that the second term in the right-hand side of 
\eqref{saddle point eq reduced model 2} is rewritten as
\begin{align}
\sum_{t=-\infty}^{\infty}
\dashint dy \frac{x-y}{t^2+(x-y)^2} \hat{\rho}(y)
=  \ \pi \dashint dy \coth \{ \pi(x-y)\} \hat{\rho}(y) 
\end{align}
while the third term is
\begin{align}
&\sum_{t=-\infty}^{\infty} \sum_{J=\frac{|t|}{2}}^{\infty}
\int dy \left\{
\frac{x-y}{(2J+\frac{3}{2})^2+(x-y)^2}-\frac{x-y}{(2J+\frac{1}{2})^2+(x-y)^2} \right\}
\hat{\tilde{\rho}}(y) \n
&=\sum_{J=0,\frac{1}{2},1,\cdots}^{\infty}(2J+1)
\int dy \left\{
\frac{x-y}{(2J+\frac{3}{2})^2+(x-y)^2}-\frac{x-y}{(2J+\frac{1}{2})^2+(x-y)^2} \right\}
\hat{\tilde{\rho}}(y) \n
&=-\sum_{n=0}^{\infty}\int dy \frac{x-y}{(n+\frac{1}{2})^2+(x-y)^2}\hat{\tilde{\rho}}(y) \n
&=-\frac{\pi}{2}\int dy \tanh \{ \pi(x-y)\} \hat{\tilde{\rho}}(y).
\end{align}
Then \eqref{saddle point eq reduced model 2} becomes
\begin{align}
0=&\frac{1}{t_1}x
-\pi \dashint dy \coth \{ \pi(x-y)\} \hat{\rho}(y) 
+\frac{\pi N_2}{N_1}\int dy \tanh \{ \pi(x-y)\} \hat{\tilde{\rho}}(y).
\end{align}
This is nothing but \eqref{saddle point eq} under the identification $t_1=\lambda_1$
and $t_2=\lambda_2$,
and thus $\hat{\rho}=\rho$ and $\hat{\tilde{\rho}}=\tilde{\rho}$ follow.

\subsection{Contributions from $\rho^{[t]}$ and $\tilde \rho ^{[t]}$ with $|t-s|\geq \ln \Lambda$ 
\label{contributions from rhot}}

Here, we evaluate the contributions from $\rho^{[t]}$ and $\tilde \rho ^{[t]}$ 
with  $|t-s|\geq \ln \Lambda$ in the first equation of \eqref{saddle point eq reduced model}.
We show that they are negligible 
in the $\Lambda\rightarrow \infty$ limit.
The same evaluation can be applied to the second equation of \eqref{saddle point eq reduced model}.

We assume that $\rho^{[t]}$ and $\tilde{\rho}^{[t]}$ 
have finite supports, and so there exists a region $[a,b]$ which contains all the supports.
There also exist two constants $c$ and $\tilde{c}$ satisfying
 $\rho^{[t]}(x)\leq c$ and $\tilde{\rho}^{[t]}(x)\leq \tilde{c}$, respectively,
for arbitrary $x\in \mathbb{R}$ and $t$.

First we evaluate such contributions in the second term in \eqref{saddle point eq reduced model},
\begin{align}
&\left(\sum_{t=s+\ln\Lambda}^{\Lambda/2}+\sum_{t=-\Lambda/2}^{s-\ln \Lambda}\right)
\int_a^b dy \frac{x-y}{(s-t)^2+(x-y)^2}\rho^{[t]}(y) \n
&\leq
 c \left(\sum_{t=s+\ln\Lambda}^{\Lambda/2}+\sum_{t=-\Lambda/2}^{s-\ln \Lambda}\right)
  \int_a^b dy \frac{x-y}{(s-t)^2+(x-y)^2} \n
&\simeq c
 \left(\sum_{n=\ln \Lambda}^{\Lambda/2-s}+\sum_{n=-\Lambda/2-s}^{-\ln\Lambda}\right)
 \left\{\int_a^b dy \frac{(x-y)}{n^2} +{\cal O}(n^{-4})\right\} \n
&\rightarrow 0 \quad (\Lambda\rightarrow \infty).
\end{align}
In the same way, those in the third term are evaluated as
\begin{align}
&\left(\sum_{t=s+\ln\Lambda}^{\Lambda/2}+\sum_{t=-\Lambda/2}^{s-\ln \Lambda}\right)
\sum_{J=\frac{|s-t|}{2}}^{\infty}
\int_a^b dy \left\{
\frac{x-y}{(2J+\frac{3}{2})^2+(x-y)^2}-\frac{x-y}{(2J+\frac{1}{2})^2+(x-y)^2} \right\}
\tilde{\rho}^{[t]}(y) \n
&\leq  
\tilde{c} \left(\sum_{t=s+\ln\Lambda}^{\Lambda/2}+\sum_{t=-\Lambda/2}^{s-\ln \Lambda}\right)
\sum_{J=\frac{|s-t|}{2}}^{\infty}
\int_a^b dy \left\{
\frac{x-y}{(2J+\frac{3}{2})^2+(x-y)^2}-\frac{x-y}{(2J+\frac{1}{2})^2+(x-y)^2} \right\} \n
&= 
\tilde{c} \left(\sum_{n=\ln \Lambda}^{\Lambda/2-s}+\sum_{n=-\Lambda/2-s}^{-\ln\Lambda}\right)
\sum_{J=\frac{|n|}{2}}^{\infty}
\int_a^b dy \left\{
-\frac{x-y}{J^3}+{\cal O}(J^{-4}) \right\} \n
& \rightarrow 0 \quad  (\Lambda\rightarrow \infty).
\end{align}

\end{document}